# The shooting methods to solve 3D nonlinear strings assemblies

**Florian SURMONT [a*]**

[a] Nantes Université, LS2N, UMR 6004, F-44000 Nantes, France, florian.surmont@univ-nantes.fr

[*] corresponding author

**Abstract**

This article presents an alternative approach to finite elements for modeling and analyzing 3D static mooring lines using string theory and the shooting method (SM) to solve two-point boundary value problems (TPBVPs) for 3D nonlinear static string equations with various boundary condition (BC) types relevant to offshore slender system assemblies.

The two-point boundary value problem for nonlinear extensible elastic strings was formulated by incorporating arbitrary 3D external distributed loads that are not restricted to gravity alone. The TPBVP was reformulated based on the formalism of the shooting method. A multi-body/multi-shooting approach is proposed to handle multi-material segments and line assemblies. A formulation of the boundary conditions that allows the modeling of Dirichlet, Robin, and mixed boundary conditions representing the displacement, force, and combined force/displacement constraints is presented.

Four validation cases are presented, comparing the results to analytical solutions: (1) a single catenary segment with ball-prismatic joint boundary conditions under several imposed forces, (2) a review of all possible boundary conditions for strings, including spring-based BCs, (3) the Velaria problem with nonlinear radial distributed load, and (4) a three-segment hanging configuration with different material properties connected by a buoy.

The results demonstrate an accuracy under $10^{-9}$ in terms of absolute errors for both positions and tensions along the entire length of the mooring lines. The proposed method also provides error control through adaptive step integration. It demonstrates high accuracy in modeling complex 3D kinematics and configurations for mooring lines, while limiting the iterative problem size.

The proposed method provides an efficient alternative to discretization-based techniques for analyzing static configurations of string kinematics slender systems with various end constraints, such as mooring lines and hawsers assemblies in offshore engineering, while maintaining simplicity in approach and implementation.

## 1. Introduction

### 1.1. State of the art

Nonlinear extensible elastic string models that handle only traction along their centerline are extensively used in the offshore industry because of their numerical time efficiency compared to more detailed beam equations. Three main techniques are used to solve static string equations [1]: dynamic relaxation, discretization-based methods (lumped mass, finite difference, finite elements), linear/nonlinear springs characterizing force-movement relationships, and catenary equations. The latter is actually a derivative of the shooting method, whose usage has been previously introduced in the offshore field to analyze the three-dimensional steady-state configuration of the underwater flexible cable problem. Application cases were restricted to the resolution of the line profile and tension under gravity and current loads on a single cable linking a surface vessel to a buggy at seabed, and single-line towing configurations [2], [3], where cables are pinned at both extremities. In offshore commercial software, catenary equations are solved using a derivative of the shooting method based on the catenary algebraic equation, as shown in this study. However, it is surprising that although the shooting method has proven its superiority over discretized methods on catenaries, a full framework adapted to the offshore industry has not yet been proposed.

The present paper tends to fill this gap by developing shooting method equations for full 3D static interconnected string equations with any kinematic joint where external loads are arbitrary and not restricted to gravity alone.

Drawing a parallel with optimal control theory [4], direct methods (of the EF type) have the advantage of not requiring a priori knowledge of the solution (the final static state) to achieve convergence, allowing constraints on the state vector (via kinematic links) to be considered in a simple manner, and are numerically robust. However, these approaches are memory- and computation-time-intensive and can be imprecise owing to the discretization method used and the time step used. Indirect methods (such as the shooting method) offer the advantage of very high accuracy at low computation times (computations can be parallelized), even for large-scale problems. Furthermore, they naturally and automatically transfer the problem of error and integration step control to the solver of the initial-value problem, thereby freeing users from these critical tasks. However, they have the disadvantage of being less robust: sensitivity to initial values and small convergence range around the desired solution. The multi-shooting technique described in the following section increases the radius of convergence of the simple method. [5]. Because the shooting method generally uses Newton's algorithms, some authors have proposed modified gradient descent techniques to improve the convergence of the algorithms [6].

In fields other than offshore engineering, the shooting method [5], [7] is used in a wide range of applications, from medical robot-assisted surgery [8], [9] to periodic determination of the stator rotor assembly [10]. One of the first to investigate the use of optimal control theory and the Pontryagin Maximum Principle in solving the TPBVP was Keller [11], [12], who developed the principles of single and multiple shooting methods. Since then several shooting methods "types" can be found in the literature from the multiple shooting method [6], [13], to continuation techniques [14], as well as shooting method applied on Lie group manifolds [15].

To adapt these techniques to interconnected strings, this paper aims to make use of classical single and multiple shooting methods to solve the statics of strings, with a special emphasis on their capability to consider several types of boundary conditions often present in the offshore industry.

First, general 3D TPBVP static equations based on [16] are presented for strings kinematics, with the introduction of arbitrary 3D external distributed loads. Particular attention is paid to the description of different boundary condition types, namely Dirichlet, Robin, and mixed boundary conditions. In the context of studying strings, these types of boundary conditions correspond to spherical, prismatic, planar joints, imposed forces at the tip, or stiffeners, which are often used in offshore field models.

Subsequently, the shooting method principle is presented to solve the problem of the structural mechanics of strings, where the TPBVP is solved iteratively as a succession of initial value problems (IVPs). The proposed motion equation state variables formalism, associated with the shooting method formalism, allows us to write naturally the mentioned boundary conditions to be able to impose a pure force, a pure displacement, or a mix force/displacement in the same way. We will then explore the single and multiple shooting methods to solve these problems. Finally, this study proposes a new kind of shooting method, namely the multi-body/multi-shooting method, to solve the structural mechanics of static strings.

The proposed formalism is validated against several numerical experiments to investigate how the proposed shooting method can solve the TPBVP of static strings. Four validation cases are proposed to extensively test the precision of the method, its capacity to handle different boundary condition types, its capacity to handle nonlinear (following) lineic loads, and its capacity to simulate assemblies. Each case is compared to a semi-analytic formulation based on algebraic catenary closed-form equations.

### 1.2. The shooting method

The shooting method was originally developed to solve Two-Point Boundary Value Problems (TPBVP) governed by an ordinary differential equation (ODE) with initial and final conditions [11]. Based on Pontryagin's maximum principle [17] from optimal control, it determines the target solutions for differential-time problems [4]. It is an indirect method, distinct from direct methods, such as finite elements and finite differences, which discretize the problem. Owing to its accuracy, it has been used for orbital problems [4] to compute the control laws for the target trajectories. In mechanics, the shooting method has been used to model articulated robots [18] and continuous flexible robots in the medical field [8], offering real-time computational advantages over discrete methods.

To illustrate this method, let us consider solving a boundary problem based on the following second-order ODE:

$$\begin{gathered} y''(x) = F\big(x, y(x), y'(x)\big),\ x \in [a, b] \\ y(a) = \alpha, y(b) = \beta \end{gathered} \tag{1}$$

A solution to this type of problem, involving Dirichlet edge conditions, does not necessarily exist and is not necessarily unique. On the other hand, the initial value problem (IVP) (2) exhibiting Neumann boundary condition type with the same differential equation typically has a unique solution that depends on the chosen value in condition $y'(a) = t$.

$$\begin{gathered} y''(x) = F\big(x, y(x), y'(x)\big), x \in [a, b] \\ y(a) = \alpha, y'(a) = t \end{gathered} \tag{2}$$

One method to solve (1) is to find $t = \bar{t}$ such that the corresponding solution $y = y(x, \bar{t})$ of the problem with initial conditions (2) satisfies the boundary condition $y(b) = y(b, \bar{t}) = \beta$. This is equivalent to search $t$ that cancels the constraint function $C(t)$:

$$C(t) = y(b, t) - \beta \tag{3}$$

Using (2) and (3), the family of methodologies for addressing boundary value problems, which entails solving a sequence of initial value problems (4), is called the shooting methods (SM):

$$SM: \begin{cases} IVP: \begin{cases} y''(x) = F(y, y', x) \\ y(x = a) = \alpha \\ y'(x = a) = t \end{cases} \\ C(t) = y(b, t) - \beta = 0 \end{cases} \tag{4}$$

In the general nonlinear case described by (4), the solution cannot be expressed as a linear combination of the solutions of two initial-value problems, and the solution of the nonlinear problem is obtained from a sequence of solutions of problems with initial values involving the slope $t = t_k$ such that:

$$\lim_{k \to \infty} y(b, t_k) = y(b, \bar{t}) = y(b) = \beta \tag{5}$$

The procedure is to adjust the slope in the same manner as when aiming at a static target; hence, the name of the method (Figure 1). Subsequently, the procedure consists in initiating shooting at slope $t_0$ from point $(a, \alpha)$. If $y(b, t_0)$ is not sufficiently close to $\beta$, the elevation is successively corrected at $t_1$, $t_2$ until $y(b, t_k)$ is close enough to touch $\beta$ (Figure 2).

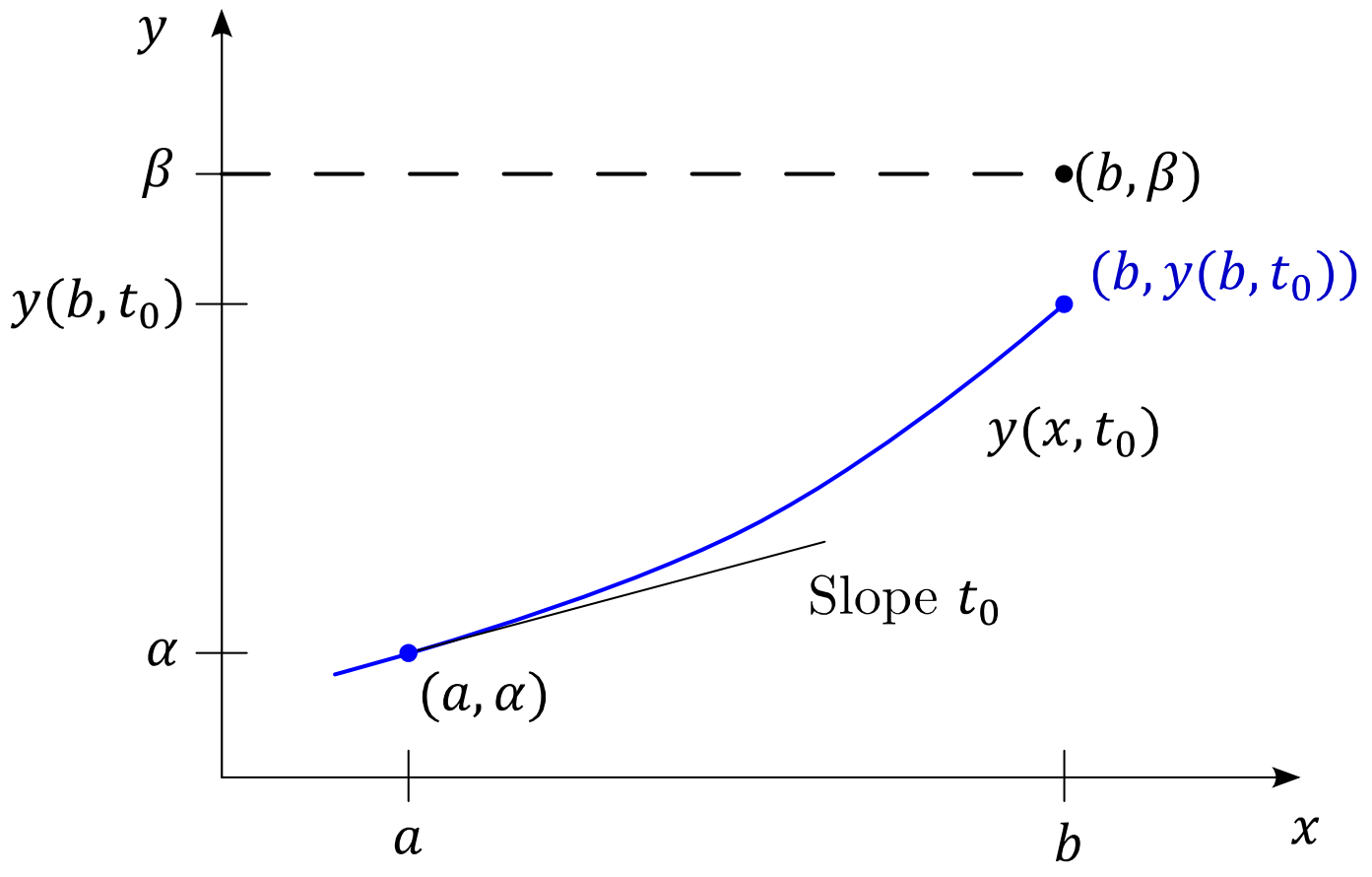

**Figure 1 Shooting method illustration**

The sequence of slopes $t_k$ is obtained by solving the constraint function described in (3). Newton's method is the most widely used, and the actualization of $t_k$ is:

$$t_{k+1} = t_k - \frac{C(t_k)}{\frac{dC}{dt}(t_k)} = t_k - \frac{y(b, t_k) - \beta}{\frac{dy}{dt}(b, t_k)} \tag{6}$$

The term $\frac{dy}{dt}(b, t_k)$ is not known and can be evaluated numerically using a finite difference.

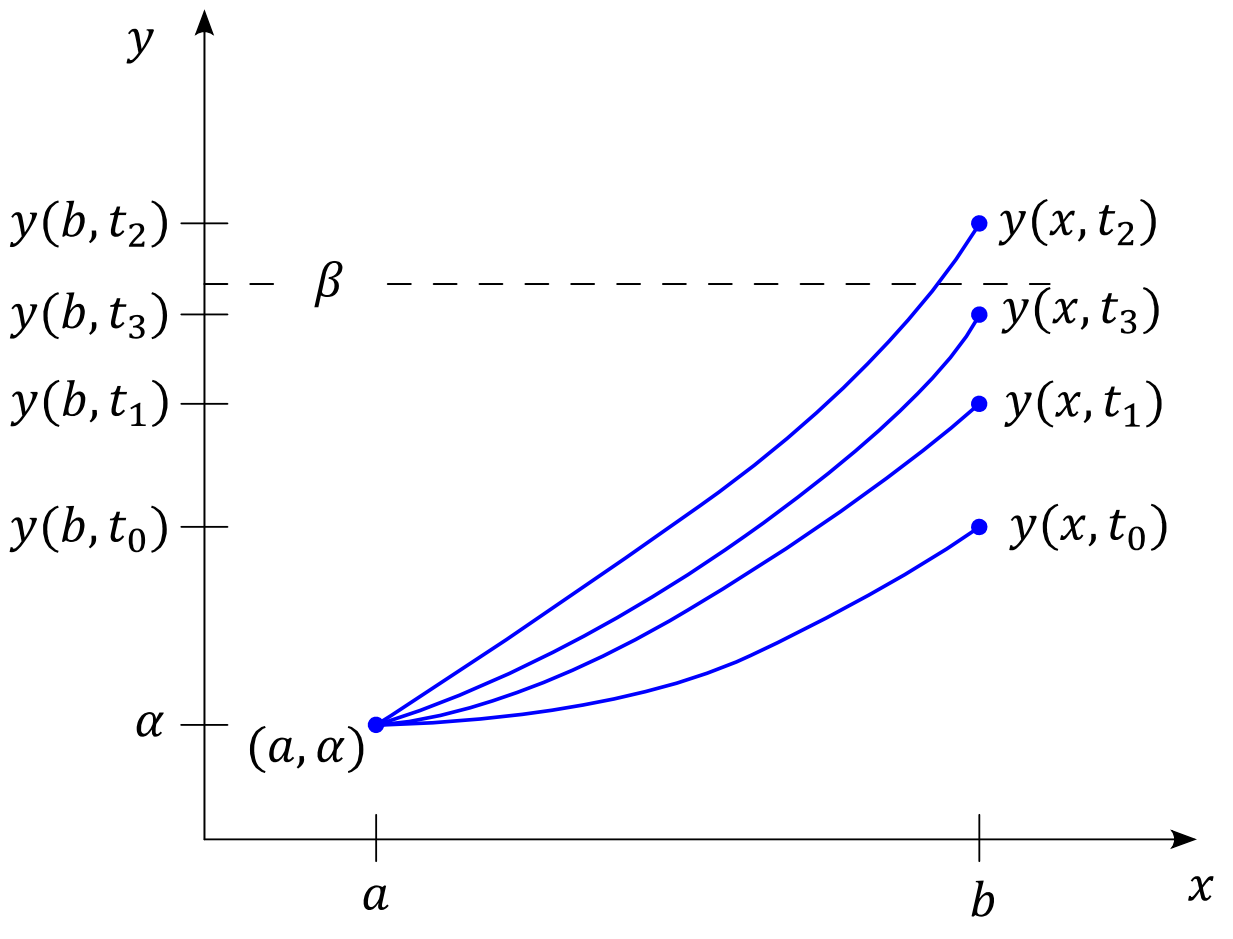

**Figure 2 Shooting method iterations**

## 2. The TPBVP of static strings

This section examines the differential equations of static strings (unconstrained or lying on a surface), various constitutive laws (elastic or undeformable), and the formulation of equilibrium equations. The boundary conditions as kinematic joints are remembered, and how they are seen in terms of the proposed formulation.

The boundary problem of a string model consists of an ordinary differential equation associated with two boundary conditions at both ends of the segment. The string theory formulation derives from the formalism defined in [16]. No assumptions of small deformations or displacements are made, and the string can be subjected to large deformations in extension. The assumptions made for the kinematics of the string are as follows:

(i) the string configuration is defined as a set of material points in space (material sections) representing a curve $\boldsymbol{r}(s) \in \mathbb{R}^3$; parameterized by arc length $s$.

(ii) the string is assumed to be perfectly flexible and subject only to extensional deformations along the neutral fiber;

(iii) the functions defining the geometric and mechanical properties are sufficiently regular to be derived as many times as necessary (generally a class $C^1$ or $C^2$ is sufficient).

Thus, from a mechanical perspective, there are no assumptions beyond the general kinematics of strings (ii). The next parts of this section only remind the mathematics around this representation.

### 2.1. Kinematics of the material point

From (i), a material point is parameterized by describing a global vector $\boldsymbol{r}(s)$ identified in a global coordinate system using the parameter $s$. This parameter is taken as the curvilinear abscissa on the reference configuration (unconstrained configuration) $\boldsymbol{r}^0(s)$ such that $|\partial_s \boldsymbol{r}^0(s)| = 1$ where $\partial_s \cdot$ denotes the partial derivative with respect to parameter $s$.

From (i) and (ii), elongation $v$ is defined as $v(s) = |\partial_s \boldsymbol{r}(s)| > 0$. On the reference configuration, the curvilinear abscissa is defined on the interval $s \in [0, L]$. The undeformed length is obtained by the relation $L = \int_0^L v^0(s) ds$. The deformed length is given by $l = \int_0^L v(s) ds$.

In this geometric description, we introduce the elongation vector $\boldsymbol{v}(s)$ which represents the deformations along the neutral fiber of the deformed configuration, and its norm $v(s)$ which is the elongation value. Then comes the differential equation describing the geometric deformations of the string statics problem:

$$\partial_s \boldsymbol{r}(s) = \boldsymbol{v}(s) = v(s) \frac{\partial_s \boldsymbol{r}(s)}{|\partial_s \boldsymbol{r}(s)|} \tag{7}$$

### 2.1. Constitutive law

The constitutive law characterizes the string's material properties through a relationship between internal forces $\boldsymbol{n}$ with the change in string shape for any configuration $\boldsymbol{r}$. Assumption (ii) imposes that internal forces and strain vector to be colinear. This leads to the constraint $\partial_s \boldsymbol{r}(s) \times \boldsymbol{n}(s) = \boldsymbol{0}$, $\forall s \in [0, L]$. Using (7) we then define the constitutive law as:

$$\boldsymbol{v}(s) = \hat{v}(|\boldsymbol{n}(s)|, s) \frac{\boldsymbol{n}(s)}{|\boldsymbol{n}(s)|} \tag{8}$$

where $\hat{v}(|\boldsymbol{n}(s)|, s) = v(s) > 0$ is the elongation, which depends only on the stress vector norm. This definition obviously requires $|\boldsymbol{n}(s)| \neq 0$, which is a direct consequence of (ii).

Using the constitutive law, the geometrical equation is rewritten:

$$\partial_s \boldsymbol{r}(s) = \hat{v}(|\boldsymbol{n}(s)|, s) \frac{\boldsymbol{n}(s)}{|\boldsymbol{n}(s)|} \tag{9}$$

Equation (8) represents the general form of the constitutive law for a string. Depending on the elongation function $\hat{v}(|\boldsymbol{n}(s)|, s)$ considered, it is possible to model linear elastic, hyperelastic, or inextensible behavior. Only linear elasticity and inextensible elasticity are addressed in the following sections.

#### 2.1.1. Inextensible constitutive law

An inextensible string is characterized by a constant strain $\hat{v}(|\boldsymbol{n}(s)|, s) = 1$. This naturally leads to the following relationship:

$$\boldsymbol{v}(s) = \frac{\boldsymbol{n}(s)}{|\boldsymbol{n}(s)|} \tag{10}$$

#### 2.1.2. Linear elastic constitutive law

A linear elastic constitutive law exhibits a linear relation between $\hat{v}(|\boldsymbol{n}(s)|, s)$ and $|\boldsymbol{n}(s)|$, such that $\hat{v}(|\boldsymbol{n}(s)|, s) = 1 + \frac{|\boldsymbol{n}(s)|}{E(s)A(s)}$, which yields from (8):

$$\boldsymbol{v}(s) = \left(1 + \frac{|\boldsymbol{n}(s)|}{EA}\right) \frac{\boldsymbol{n}(s)}{|\boldsymbol{n}(s)|} \tag{11}$$

where $E(s)$ and $A(s)$ are respectively Young's modulus and material cross-sectional area, which may depend on the curvilinear abscissa.

### 2.2. Static equilibrium equations

Forces acting on a generic segment $(a, s)$ with $0 < a < s < L$ are the internal forces $\boldsymbol{n}^+(s)$ exerted by the segment $(s, L)$ on $(a, s)$ and $-\boldsymbol{n}^-(a)$ are the internal forces exerted by the segment $(0, a)$ on segment $(a, s)$.

We assume that all other types of (external) forces applied to a segment are of the form $\int_a^s \boldsymbol{f}(\xi) d\xi$ where $\boldsymbol{f}(s)$ are the forces per unit length (also called *distributed forces*) acting on the string. Vectors $\boldsymbol{n}^+$, $\boldsymbol{n}^+$ and $\boldsymbol{f}$ are defined in the global coordinate system. For any $0 < a < s < L$, the static equilibrium equation of the segment $(a, s)$ can be written as:

$$\boldsymbol{n}^{+}(s) - \boldsymbol{n}^{-}(a) + \int_{a}^{s} \boldsymbol{f}(\xi) d\xi = 0 \tag{12}$$

Continuity condition (iii) requires $\boldsymbol{n}^{+}(a) = \boldsymbol{n}^{-}(a) = \boldsymbol{n}(a)$ when $s \to a$. Assuming an elementary segment of length $ds$then the sign convention for $\boldsymbol{n}$ is as follows: the forces exerted by the material point at $s + ds$ on the material point located at $s$ (from right to left) are positive.

Finally, by differentiating (12) with respect to the curvilinear abscissa $s$, the problem's second differential equation for the tension vector $\boldsymbol{n}$ is obtained:

$$\partial_s \boldsymbol{n}(s) = -\boldsymbol{f}(s) \tag{13}$$

### 2.3. Boundary conditions

This section discusses the boundary conditions for strings, which are essential for solving differential equations at domain boundary points. There are four primary types: Cauchy, Dirichlet, Neumann, and Robin. Cauchy conditions provide values for state variables and their derivatives, akin to solving an initial value problem. Dirichlet conditions specify the state variable values, Neumann conditions focus on their derivatives, and Robin conditions involve a linear combination of both. The proposed resolution method is applicable to all boundary condition types, without differentiation.

In a second order differential problem where $\mathrm{y}''(\mathrm{x}) = \mathrm{F}(\mathrm{x}, \mathrm{y}(\mathrm{x}), \mathrm{y}'(\mathrm{x}))$, the state variable refers usually to $\mathrm{y}(x)$ which corresponds in our case to the position field $\boldsymbol{r}(s)$. In our case, the derivative of the state $\mathrm{y}'(\mathrm{x})$ refers to the vector $\partial_s \boldsymbol{r}$, which is assimilated to the internal force field $\boldsymbol{n}(s)$ through the constitutive laws. Formally, the boundary condition of a second order differential problem constrains a couple of $\mathrm{y}(\mathrm{x})$ and/or $\mathrm{y}'(\mathrm{x})$ at the boundaries of the integration interval. In mechanics, this leads to constraints on fields $\boldsymbol{r}$ and/or $\boldsymbol{n}$ at $s = 0$ and $s = L$, imposed by the type of kinematic joint used at the extremity of the string.

Table 1 defines the boundary conditions for permissible kinematic joints in strings. Due to the absence of rotational kinematics (infinite flexibility or zero stiffness), certain kinematic connections - clamp, revolute, and spherical joint - share the same free and constrained degrees of freedom. It is interesting to note that a kinematic joint always constrains half of the degrees of freedom of the gathered fields $\boldsymbol{r}$ and $\boldsymbol{n}$, i.e., 3 variables are always constrained, whereas 3 others are always remained unknown, which is totally expected because the string problem is a TPBVP.

| | Connection type | Free DoFs $\boldsymbol{X}$ | Constrained DoFs $\boldsymbol{Y}$ |
|---|---|---|---|
| (a) | Clamp<br>Revolute<br>Spherical | $\boldsymbol{n}$ | $\boldsymbol{r}$ |
| (b) | Imposed force | $\boldsymbol{r}$ | $\boldsymbol{n}$ |
| (c) | Prismatic of axis $\boldsymbol{x}$<br>Cylindrical of axis $\boldsymbol{x}$<br>Screw of axis $\boldsymbol{x}$<br>Sphere-cylinder of axis $\boldsymbol{x}$ | $x \quad n_y \quad n_z$ | $y \quad z \quad n_x$ |
| (d) | Sphere-plan of axis $\boldsymbol{z}$<br>Cylinder-plan of normal $\boldsymbol{z}$<br>Plan-plan of axis $\boldsymbol{z}$ | $x \quad y \quad n_z$ | $z \quad n_x \quad n_y$ |
| (e) | Linear spring | $x \quad y \quad z$ | $n_x(x,y,z) \quad n_y(x,y,z) \quad n_z(x,$ |
| (f) | Linear spring of axis $\boldsymbol{x}$ | $x \quad n_y \quad n_z$ | $n_x(x) \quad y \quad z$ |
| ⋮ | ⋮ | ⋮ | ⋮ |

**Table 1 Boundary conditions for strings kinematics: unknown and constrained fields**

### 2.4. Summary of the TPBVP equations

In previous sections, we reviewed the mathematical principles underlying the mechanical representation of general strings. This section aims to extract uniform notations for the TPBVP equations to solve them directly, without simplification, using the shooting method.

#### 2.4.1. Differential equation

We introduce the vector $\boldsymbol{\varphi}$ such as $\boldsymbol{\varphi}(s) = [\boldsymbol{r}(s), \boldsymbol{n}(s)]^T$ to put the differential equations in the form:

$$\partial_s \boldsymbol{\varphi} = F(\boldsymbol{\varphi}, s) \tag{14}$$

Note that this formalism is compatible with both unconstrained strings and strings lying on a surface. Indeed, differential equations (9) and (13) for unconstrained string yield:

$$\partial_s \boldsymbol{\varphi} = \begin{cases} \partial_s \boldsymbol{r} = \dfrac{\hat{v}(|\boldsymbol{n}|)}{|\boldsymbol{n}|} \boldsymbol{n} \\ \partial_s \boldsymbol{n} = -\boldsymbol{f} \end{cases} \tag{15}$$

#### 2.4.2. Constraint equations

As seen in Section 2.3, a kinematic joint imposes half of the variables of the vector $\boldsymbol{\varphi}$. We note respectively $\boldsymbol{X}$ and $\boldsymbol{Y}$ respectively these free and constrained fields, which are components of $\boldsymbol{\varphi}$, and whose union reconstitutes $\boldsymbol{\varphi}$ at $s = 0$ and $s = L$. These $\boldsymbol{X}$ and $\boldsymbol{Y}$ fields can indifferently represent positions, forces, or their combination, as mentioned in Table 1. With this notation, the kinematic joint constraint is written in a general form:

$$\boldsymbol{C}(\boldsymbol{Y}, \overline{\boldsymbol{Y}}) = \boldsymbol{0} \tag{16}$$

where $\overline{\boldsymbol{Y}}$ denotes the evaluation of the variable $\boldsymbol{Y}$, imposed by the kinematic joint. As an example, a ball joint positioned at $\overline{\boldsymbol{r}} = (\bar{x}, \bar{y}, \bar{z})$ would constrain the position vector $\boldsymbol{r}$ and yield the constraint equation $\boldsymbol{C}(\boldsymbol{Y} = \boldsymbol{r}, \overline{\boldsymbol{Y}} = \overline{\boldsymbol{r}}) = \boldsymbol{Y} - \overline{\boldsymbol{Y}} = \boldsymbol{r} - \overline{\boldsymbol{r}} = \boldsymbol{0}$

We define the functions $\boldsymbol{\Gamma}_{\boldsymbol{X}}$ and $\boldsymbol{\Gamma}_{\boldsymbol{Y}}$ as maps from the state vector of the differential problem to the free and constrained fields of the kinematic links.

$$\boldsymbol{X} = \boldsymbol{\Gamma}_{\boldsymbol{X}}(\boldsymbol{\varphi}) \tag{17}$$

$$\boldsymbol{Y} = \boldsymbol{\Gamma}_{\boldsymbol{Y}}(\boldsymbol{\varphi}) \tag{18}$$

We also introduce the inverse function $\boldsymbol{\gamma}$ that maps constrained fields $\boldsymbol{Y}$ and unconstrained fields $\boldsymbol{X}$ to $\boldsymbol{\varphi}$.

$$\boldsymbol{\varphi} = \boldsymbol{\gamma}(\boldsymbol{X}, \boldsymbol{Y}) \tag{19}$$

$\boldsymbol{\gamma}$ allows us to move from the kinematically admissible variables of a boundary condition to the state vector of the differential problem.

In the next, subscripts $\cdot_{0,L}$ are used to locate indifferently $\boldsymbol{X}$, $\boldsymbol{Y}$, $\overline{\boldsymbol{Y}}$, $\boldsymbol{\varphi}$, $\boldsymbol{C}$, $\boldsymbol{\gamma}$, $\boldsymbol{\Gamma}$ fields or functions at $s = 0$ and $s = L$. As an example, vectors $\boldsymbol{\varphi}_0$ and $\boldsymbol{\varphi}_L$ are built upon kinematically free variables $\boldsymbol{X}_0$, $\boldsymbol{X}_L$ and constrained variables $\boldsymbol{Y}_0$, $\boldsymbol{Y}_L$ such that $\boldsymbol{\varphi}_{0,L} = \boldsymbol{\gamma}_{0,L}(\boldsymbol{X}_{0,L}, \boldsymbol{Y}_{0,L})$ where $\boldsymbol{\gamma}_0$ and $\boldsymbol{\gamma}_L$ are respectively the transition functions at $s = 0$ and $s = L$, with $\boldsymbol{C}_{0,L}(\boldsymbol{Y}_{0,L} = \boldsymbol{\Gamma}_{\boldsymbol{Y}_{0,L}}(\boldsymbol{\varphi}_{0,L}), \overline{\boldsymbol{Y}}_{0,L}) = 0$

Recalling the general formulations (14) and (16), we can rewrite the TPBVP into a first order ODE system (20) with the boundary conditions defined by (21) and (22). The formalism proposed here for the boundary conditions constraints allows us to address both free and imposed fields in position/rotation, imposed, or mixed forces/moments.

$$\partial_s \boldsymbol{\varphi}(s) = \boldsymbol{F}(\boldsymbol{\varphi}, s) \tag{20}$$

$$\boldsymbol{C}_0(\boldsymbol{Y}_0, \overline{\boldsymbol{Y}}_0) = \boldsymbol{0} \tag{21}$$

$$\boldsymbol{C}_L(\boldsymbol{Y}_L, \overline{\boldsymbol{Y}}_L) = \boldsymbol{0} \tag{22}$$

The two-points boundary value problem consists of solving the system (20) to (22), where equations (21) and (22) describe the most general case of separate boundary conditions.

For mixed boundary conditions, it is important to pay attention to the basis in which the components of positions, orientations, forces and moments are written. Mechanical connections often exhibit a preferred basis in which translational motions can be written more easily. As an example, a prismatic joint of axis $\boldsymbol{t}$ would be written as $\boldsymbol{C} = \boldsymbol{n} \cdot \boldsymbol{t} = 0$. Additionally, the constraint functions $\boldsymbol{C}_{0,L}$ may also be nonlinear.

When the fields $\boldsymbol{n}$ or $\boldsymbol{r}$ are fully constrained, the constraint functions simplify to (23). This is particularly the case for clamp, ball or force-imposed connections.

$$\boldsymbol{C}_{0,L}(\boldsymbol{Y}_{0,L}, \overline{\boldsymbol{Y}}_{0,L}) = \boldsymbol{Y}_{0,L} - \overline{\boldsymbol{Y}}_{0,L} = \boldsymbol{\Gamma}_{\boldsymbol{Y}_{0,L}}(\boldsymbol{\varphi}_{0,L}) - \overline{\boldsymbol{Y}}_{0,L} \tag{23}$$

When we set up the kinematics of strings, we defined the internal force fields $\boldsymbol{n}$ as the action of forces exerted by the section located "to the right" of the section under study (regarding the direction of the curvilinear abscissa). At the ends (edges), these forces represent either the action of the segment on the connection or the action of the connection on the segment. To enable static equilibrium of slender bodies subjected to forces imposed at the ends, we constrain the fields $\boldsymbol{n}$ at the ends such that the forces transmitted by the link (sum of internal and imposed forces) are effectively zero. Thus, an imposed force $\boldsymbol{F}$ will be seen as a free link constraining the field $\boldsymbol{n}$ such that $\boldsymbol{n} = \pm \boldsymbol{F}$. The sign $\pm$ comes from the positive convention for force $\boldsymbol{n}$, as well as the location of the boundary condition (start or end).

Finally, the spring-loaded force connections can be derived by making the force values dependent on the kinematics. Thus, to combine a prismatic axis connection $\boldsymbol{t}$ with a spring of stiffness $k$ and a reference (attachment) point $P$ coordinates $\boldsymbol{p}$**,** the constraint on the force component will be $\boldsymbol{n}.\boldsymbol{t} = n_t(\boldsymbol{r}) = \pm k(\boldsymbol{p} - \boldsymbol{r}) \cdot \boldsymbol{t}$.

## 3. Shooting methods

Now that we have written the two-points boundary value problem for strings in a canonical form in system (20) to (22), we can make use of the shooting method.

As a first step, we rewrite the two-points boundary value problem equations within the framework of the single-shooting method to develop the formalism in accordance with the previous notations.

Within this formalism, we then develop the multiple shooting equations used to model geometric and mechanical discontinuities to free ourselves from the constraint of derivability of the geometric and mechanical fields. (iii) imposed when setting up the string kinematics.

Finally, the limitations of the classical multi-shooting approach for managing interconnected segments are highlighted, and a formulation based on a multi-body approach is proposed.

### 3.1. Single shooting method

The single shooting method consists of transforming the TPBVP (20) to (22) in the search for the zeros of a (generally nonlinear) constraint function evaluated from a succession of initial value problems (IVP) (24).

$$\boldsymbol{\varphi}(s) = \int_0^s \partial_s \boldsymbol{\varphi} + \boldsymbol{\varphi}_0 \qquad (24)$$

For clarity, we first assume that $\boldsymbol{\varphi}_0$ is simply a concatenation of vectors $\boldsymbol{X}_0$ and $\boldsymbol{Y}_0$. This corresponds to separated fields boundary conditions (a), (b) of Table 1 or boundary conditions aligned with global axes $\boldsymbol{x}$, $\boldsymbol{y}$ or $\boldsymbol{z}$ (for cases (c), (d), (e), (f), etc.). We then develop the general case in which the boundary condition constraints are expressed in any arbitrary frame.

The principle of the classic single-shooting method in the notation previously introduced is to solve the constraints $\boldsymbol{C}$ of the *shooting problem*, imposed by the boundary condition at $s = L$, with respect to *guesses* $\boldsymbol{X}_0$ of the free fields of the boundary condition at $s = 0$.

The initial state vector for the IVP is built such that $\boldsymbol{\varphi}_0 = \boldsymbol{\gamma}_0\left(\boldsymbol{X}_0, \boldsymbol{Y}_0 = \overline{\boldsymbol{Y}}_0(\boldsymbol{X}_0)\right) = \boldsymbol{\gamma}_0(\boldsymbol{X}_0)$ from (19). Note that $\overline{\boldsymbol{Y}}_0$ may depend on $\boldsymbol{X}_0$ (i.e., spring or helical case).

Integrating this initial state $\boldsymbol{\varphi}_0$ through the ordinary differential equation (ODE) allows us to evaluate the final state $\boldsymbol{\varphi}(s = L)$ which then depends on the *shooting variables* $\boldsymbol{X}_0$ such that $\boldsymbol{\varphi}(s = L; \boldsymbol{\varphi}_0) = \boldsymbol{\varphi}_L(\boldsymbol{\gamma}_0(\boldsymbol{X}_0))$.

Mapping (18) let us extract the imposed variable $\boldsymbol{Y}_L(\boldsymbol{X}_0) = \boldsymbol{\Gamma}_{\boldsymbol{Y}_L}\left(\boldsymbol{\varphi}_L(\boldsymbol{\gamma}_0(\boldsymbol{X}_0))\right)$ from $\boldsymbol{\varphi}_L$. Moreover, as $\overline{\boldsymbol{Y}}_0$ may depends on $\boldsymbol{X}_0$, $\overline{\boldsymbol{Y}}_L$ may also depends on $\boldsymbol{X}_L$ leading to $\boldsymbol{X}_L(\boldsymbol{X}_0) = \boldsymbol{\Gamma}_{\boldsymbol{X}_L}\left(\boldsymbol{\varphi}_L(\boldsymbol{\gamma}_0(\boldsymbol{X}_0))\right)$ such that at $s = L$: $\overline{\boldsymbol{Y}}_L(\boldsymbol{X}_0) = \overline{\boldsymbol{Y}}_L\left(\boldsymbol{\Gamma}_{\boldsymbol{X}_L}\left(\boldsymbol{\varphi}_L(\boldsymbol{\gamma}_0(\boldsymbol{X}_0))\right)\right)$.

Altogether, the constraints $\boldsymbol{C}$ of the *shooting problem* explicitly depend on the variable $\boldsymbol{X}_0$ of the *shooting problem* such that:

$$\boldsymbol{C}(\boldsymbol{X}_0) = \boldsymbol{C}_L\left(\boldsymbol{Y}_L\left(\boldsymbol{\varphi}_L\left(\boldsymbol{\varphi}_0\left(\boldsymbol{X}_0, \overline{\boldsymbol{Y}}_0(\boldsymbol{X}_0)\right)\right)\right), \overline{\boldsymbol{Y}}_L\left(\boldsymbol{X}_L\left(\boldsymbol{\varphi}_L\left(\boldsymbol{\varphi}_0\left(\boldsymbol{X}_0, \overline{\boldsymbol{Y}}_0(\boldsymbol{X}_0)\right)\right)\right)\right)\right) \tag{25}$$

In this description, $\boldsymbol{X}_0$ are the free variables of the initial boundary condition and the constraint function $\boldsymbol{C}$ depends only on $\boldsymbol{X}_0$.

An extension of this description is to assume both $\boldsymbol{X}_0$ and $\boldsymbol{Y}_0$ as variable of the shooting problem. The resolution process is the same, except that the initial $\boldsymbol{Y}_0$ is also estimated during the shooting process. However, the constraint function $\boldsymbol{C}$ will depend on both $\boldsymbol{X}_0$ and $\boldsymbol{Y}_0$ will be replaced by (26).

$$\boldsymbol{C}(\boldsymbol{X}_0, \boldsymbol{Y}_0) = \begin{cases} \boldsymbol{C}_0\left(\boldsymbol{Y}_0, \overline{\boldsymbol{Y}}_0(\boldsymbol{X}_0)\right) \\ \boldsymbol{C}_L\left(\boldsymbol{Y}_L\left(\boldsymbol{\varphi}_L(\boldsymbol{\varphi}_0(\boldsymbol{X}_0, \boldsymbol{Y}_0))\right), \overline{\boldsymbol{Y}}_L\left(\boldsymbol{X}_L\left(\boldsymbol{\varphi}_L(\boldsymbol{\varphi}_0(\boldsymbol{X}_0, \boldsymbol{Y}_0))\right)\right)\right) \end{cases} \tag{26}$$

Regardless of the choice of approach, we can write the constraint function depending on the vector $\boldsymbol{Z}$ such as $\boldsymbol{Z}_0 = \boldsymbol{X}_0$ in the case (25) and $\boldsymbol{Z}_0 = [\boldsymbol{X}_0, \boldsymbol{Y}_0]^T$ in case (26). In addition, from (17) and (18), we can define $\boldsymbol{Z} = \boldsymbol{\Gamma}_{\boldsymbol{Z}}(\boldsymbol{\varphi})$ and $\boldsymbol{\varphi} = \boldsymbol{\gamma}(\boldsymbol{Z})$ without ambiguity.

The shooting problem is then written as the nesting of an IVP within a Root Finding (RF) algorithm:

$$RF:\begin{cases} IVP:\begin{cases}\partial_s\boldsymbol{\varphi}(s) = \boldsymbol{F}(\boldsymbol{\varphi}, s)\\ \boldsymbol{\varphi}(s=0) = \boldsymbol{\varphi}_0 = \boldsymbol{\gamma_0}(\boldsymbol{Z}_0)\end{cases}\\ \boldsymbol{C}(\boldsymbol{Z}_0) = \boldsymbol{0}\end{cases} \quad (27)$$

$\boldsymbol{C}$ is the constraint vector function and $\boldsymbol{\varphi}_0$ may be partially or totally unknown, and depends on the variable $\boldsymbol{Z}_0$ which is the unknown of the zero search algorithm (Newton). Two levels of notation must therefore be taken into account: lower-case variables $\boldsymbol{\varphi}$, $s$, $\boldsymbol{r}$, etc., refer to the integration problem, and uppercase variables $\boldsymbol{C}$, $\boldsymbol{X}$, $\boldsymbol{Y}$, $\boldsymbol{Z}$, etc., refer to the root finding problem.

As a reminder, the shooting method therefore consists of iterating over the unknown part $\boldsymbol{Z}$ of the IVP to respect the constraint $\boldsymbol{C}$ calculated from $\boldsymbol{Y}_L$ (and if necessary $\boldsymbol{Y}_0$) evaluated through successive integrations. The resolution process is illustrated in Figure 3.

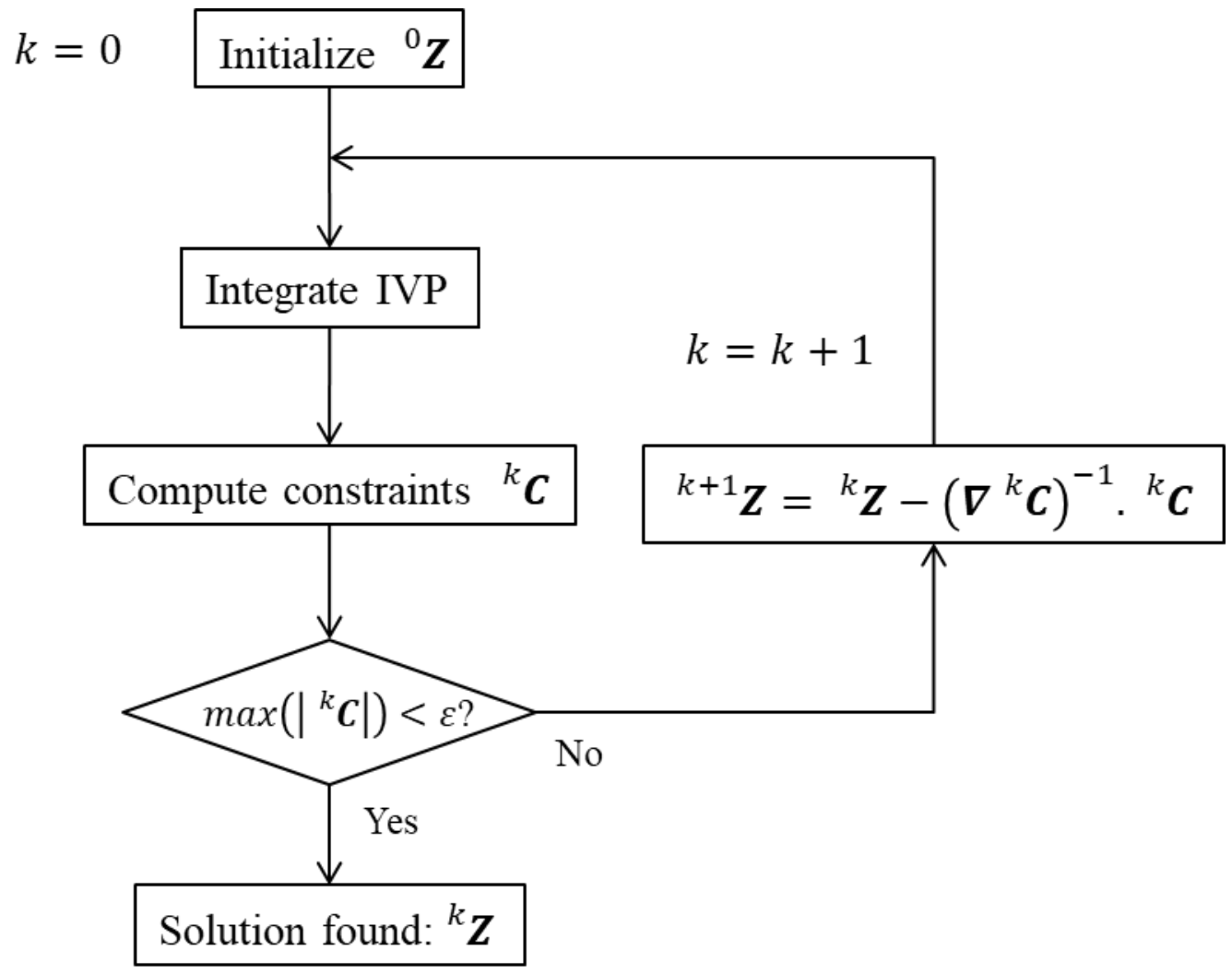

**Figure 3 Shooting method resolution process**

We used a Runge-Kutta-Fehlberg 4-5 (RKF45) scheme to integrate the ODE. This algorithm automatically controls the step size based on a specified integration error between a RK4 and a RK5 evaluations. Thus, adapting the integration step along a segment ensures that the cumulative error remains within the prescribed tolerance, thereby providing accurate integration with minimal points.

The constraint function is solved using the Newton-Raphson algorithm, in which the Jacobian of the problem is calculated using finite differences. We used a Singular Value Decomposition (SVD) to ensure the correct inversion of the Jacobian in the Newton process. Although reputedly costly, this decomposition is actually performed on small matrices: $\boldsymbol{C}$ being a vector of size 3 or 6, leads to a Jacobian of size 3x3 or 6x6. Inverting such small matrices is highly efficient compared to discrete approaches (i.e., finite elements), where the size of the Jacobian (stiffness matrix) is proportional to NxN, where N represents the number of nodes.

As we'll see later, the shooting method allows convergence in less than 10 iterations in general, number of "costly" matrix decomposition is then very limited.

From a general point of view, the shooting method, as previously described, inherits the drawbacks of a nonlinear resolution algorithm: if the function $\boldsymbol{C}$ is highly nonlinear, it may not converge if the estimate ${}^{0}\boldsymbol{Z}$ is too far from a root of $\boldsymbol{C}$. This notion of "distance" is based on the physical equations involved, the shape of $\boldsymbol{C}$ and the conditioning (eigenvalues) of $\nabla\boldsymbol{C}$. Stiff problems - where small perturbations of $\mathbf{Z}$ induce large variations in $\boldsymbol{C}$ through the ODE - restrict the convergence radius of the Newton's algorithm, and require ${}^{0}\boldsymbol{Z}$ to be close to the desired solution. In addition, since $\boldsymbol{C}$ is nonlinear, several solutions may exist, and the solution obtained by the Newton's algorithm will depend on the initial iteration ${}^{0}\boldsymbol{Z}$. This behavior will be illustrated in the third numerical experiment of this study.

The convergence problems inherent in the single-shooting method are partially solved using the multiple-shooting method [5], [19]. Other methods based on modifying the slope of Newton's algorithm are also referenced in the state of the art [6].

In contrast to discretized approaches (i.e., finite elements), a significant advantage of the shooting method is its ability to handle all integrated fields and parameters in a continuous manner. For example, the elastic constitutive law introduced in (11) can depend on the curvilinear abscissa $s$ through the parameters $A(s)$ and $E(s)$, characterizing the cross-section. This enables the modeling of continuous variations in the shape of the material section of the slender body.

In the description developed herein, the unknowns $\boldsymbol{Z}_0$ thus $\boldsymbol{\varphi}_0$ depend on the boundary condition at $s = 0$. As a reminder, Table 1 (page 10) summarizes the known and unknown components for each type of kinematic link. It should be noted that the notations adopted here are not dependent on the configuration studied (clamp-clamp, clamp-ball, etc.) and allow for genericity in the treatment of the configurations. Furthermore, for a given configuration of boundary conditions, the results obtained using the shooting method are independent of the integration direction. This property will be verified in subsequent examples.

### 3.2. Multiple shooting method

The multi-shooting method [20] can be used to stabilize the simple shooting problem when the Jacobian $\nabla\boldsymbol{C}$ is poorly conditioned [19]. The principle is to split the segment into $n$ sub-segments following the same differential equation and introduce continuity equations (constraints) of the fields at cuts. This approach results in smaller integration intervals, which tend to enhance the conditioning of $\nabla\boldsymbol{C}$ at the expense of increasing the size of the Jacobian to be inverted. Integration and constraints expressed on smaller sub-intervals can be seen as linearization technique of the constraint function $\boldsymbol{C}$. Shorter integration intervals lead to smaller variations in the propagated vector $\boldsymbol{\varphi}_{s_n}$and, therefore, of the constraint function $\boldsymbol{C}$.

We propose here another application of the multi-shooting technique to model discontinuities that can appear in the fields $\boldsymbol{r}(s)$, $\boldsymbol{n}(s)$ or $\boldsymbol{f}(s)$. Indeed, assumption (iii) from string kinematics imposes the continuity of these fields. These discontinuities can be a punctual force at a given curvilinear abscissa, a sudden change in the material properties of the section (different section geometries, different material parameters), or a discontinuity in external forces (external force applied to a section only).

Regardless of the application framework, the principle of the multi-shooting method for solving a problem at both ends is based on the same idea of solving a problem at initial values but on sub-intervals of the initial domain. The multi-shooting method is therefore based on the same single shooting method system (27) but differs the sizes of $\boldsymbol{\varphi}$, $\boldsymbol{Z}$ and $\boldsymbol{C}$. The division into sub-intervals generates new unknowns and new constraints for matching the trajectories of the integrated fields at the boundaries of each sub-domain. Figure 4 illustrates the subdivision for the choice of curvilinear abscissas $s_0 = 0 < s_1 < s_i < s_{i+1} < s_n = L$.

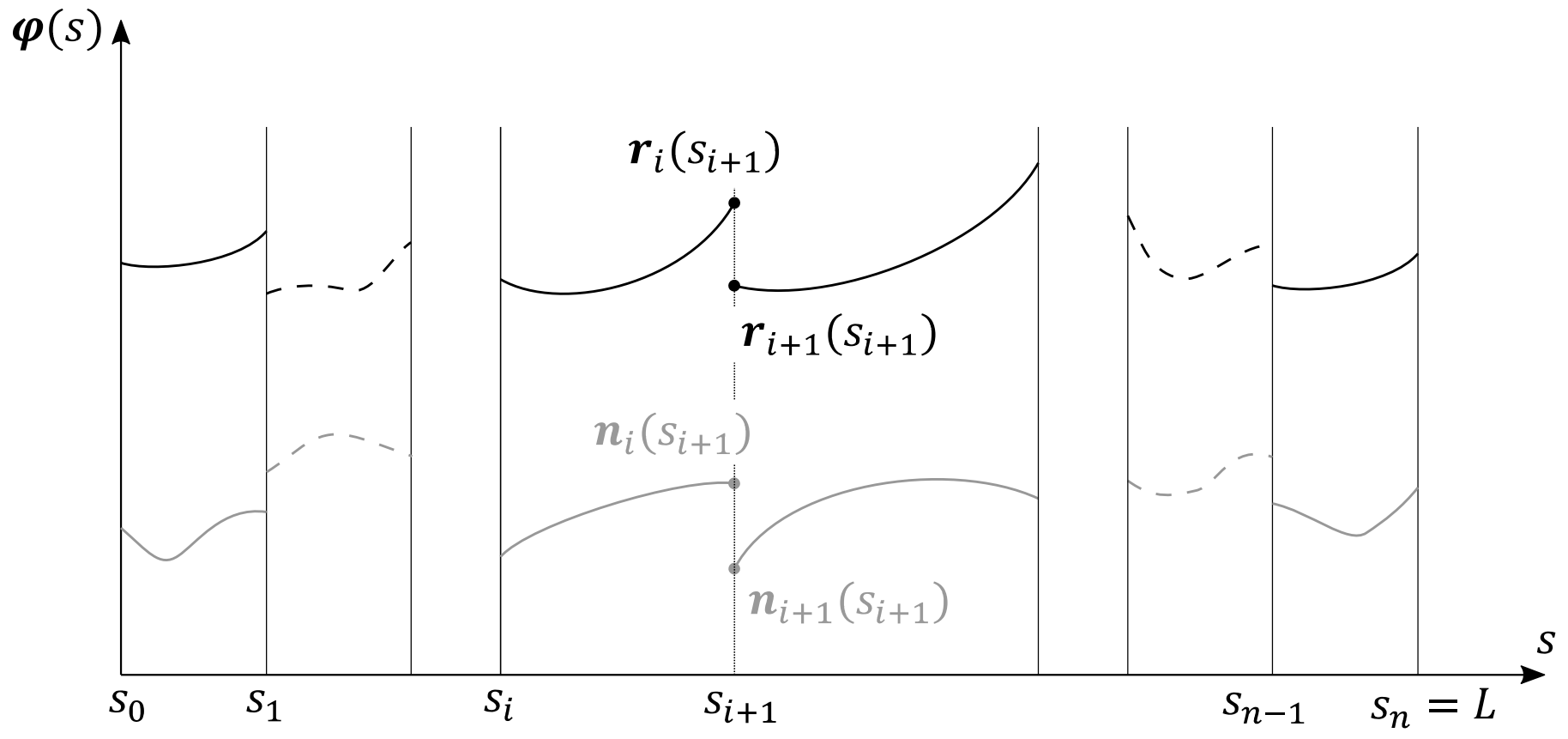

**Figure 4 Illustration of the multi-shooting method**

The solution on the $i\ th$ segment is denoted $\boldsymbol{\varphi}_{\mathrm{sb}_i}(s)$. For $0 < i < n$, the entire state vector $\boldsymbol{\varphi}_{\mathrm{sb}_i}(s = s_i) = \boldsymbol{\varphi}_{s_i}^{\mathrm{sb}_i}$ is unknown at the start of the segment $\mathrm{sb}_i$, and $\boldsymbol{\varphi}_{s_i}^{\mathrm{sb}_i}$ are additional variables of the optimization problem such that:

$$\boldsymbol{\varphi}_{\mathrm{sb}_i}(s = s_i) = \boldsymbol{\varphi}_{s_i}^{\mathrm{sb}_i} = \boldsymbol{Z}_{s_i} \tag{28}$$

We use the superscript to designate the segment number, and the lower script to designate the curvilinear abscissa location such that $\boldsymbol{\varphi}_{s_i}^{\mathrm{sb}_i} = \boldsymbol{\varphi}_{\mathrm{sb}_i}(s = s_i)$. The lower script notation for $\boldsymbol{Z}$ still represents the curvilinear abscissa location.

Note that (28) is actually a specific form of the mapping $\boldsymbol{\gamma}$ introduced in (19), such that $\boldsymbol{\gamma}_{s_i}$ is the identity operator, and (28) is expressed as $\boldsymbol{\varphi}_{s_i}^{\mathrm{sb}_i} = \boldsymbol{\gamma}_{s_i}(\boldsymbol{Z}_{s_i}) = \boldsymbol{Z}_{s_i}$.

In addition, we must ensure fields continuity at $s = s_i$, that is, the integrated fields at the end of the interval $\mathrm{sb}_{i-1}$ must be equal to the field's values at the start of the segment $\mathrm{sb}_i$. Therefore, the constraint vector $\boldsymbol{C}$ vector is augmented with the new constraint $\boldsymbol{C}_{s_i} = \boldsymbol{\varphi}_{\mathrm{sb}_{i-1}}(s = s_i) - \boldsymbol{\varphi}_{\mathrm{sb}_i}(s = s_i) = \boldsymbol{\varphi}_{s_i}^{\mathrm{sb}_{i-1}} - \boldsymbol{\varphi}_{s_i}^{\mathrm{sb}_i} = \boldsymbol{0}$.

Using mappings $\boldsymbol{\Gamma}$ and $\boldsymbol{\gamma}$ introduced in (17) and (19), we can write $\boldsymbol{C}_{s_{i+1}}$ with regard to the shooting variables in the generic form $\boldsymbol{C}_{s_i}\left(\boldsymbol{\Gamma}_{\boldsymbol{Z}_{s_i}}\left(\boldsymbol{\varphi}_{s_i}^{\mathrm{sb}_{i-1}}(\boldsymbol{Z}_{s_{i-1}})\right), \boldsymbol{\gamma}_{s_i}(\boldsymbol{Z}_{s_i})\right) = \boldsymbol{0}$. This generic form makes the methods comparable between the previously introduced single shooting method and the multi-body/multi-shooting method, which we will develop in the following sections.

Note that $\boldsymbol{\gamma}_{s_i}$ and $\boldsymbol{\Gamma}_{\boldsymbol{Z}_{s_i}}$ are identity functions such that $\boldsymbol{\varphi}_{s_i}^{\mathrm{sb}_{i-1}} = \boldsymbol{\Gamma}_{\boldsymbol{Z}_{s_i}}\left(\boldsymbol{\varphi}_{s_i}^{\mathrm{sb}_{i-1}}\right)$ and $\boldsymbol{\gamma}_{s_i}(\boldsymbol{Z}_{s_i}) = \boldsymbol{Z}_{s_i}\ \forall i \in \,]0,n[$.

Finally, by introducing vector $\boldsymbol{\phi}(s) = \left[\boldsymbol{\varphi}_{\mathrm{sb}_0}, \cdots, \boldsymbol{\varphi}_{\mathrm{sb}_i}, \cdots, \boldsymbol{\varphi}_{\mathrm{sb}_{n-1}}\right]^T$, equation (27) is rewritten as (30). Note that (30) has exactly the same structure as (27).

$$\boldsymbol{G}(\boldsymbol{\phi}, s) = \begin{bmatrix} \boldsymbol{F}(\boldsymbol{\varphi}_{\mathrm{sb}_0}, s) \\ \vdots \\ \boldsymbol{F}(\boldsymbol{\varphi}_{\mathrm{sb}_i}, s) \\ \vdots \\ \boldsymbol{F}(\boldsymbol{\varphi}_{\mathrm{sb}_{n-1}}, s) \end{bmatrix} \tag{29}$$

$$RF: \begin{cases} IVP: \begin{cases} \partial_s \boldsymbol{\phi}(s) = \boldsymbol{G}(\boldsymbol{\phi}, s) \\ \boldsymbol{\phi}_{s_0} = \left[\boldsymbol{\gamma}_{s_0}(\boldsymbol{Z}_{s_0}) \quad \cdots \quad \boldsymbol{\gamma}_{s_i}(\boldsymbol{Z}_{s_i}) \quad \cdots \quad \boldsymbol{\gamma}_{s_{n-1}}(\boldsymbol{Z}_{s_{n-1}})\right]^T \end{cases} \\ \boldsymbol{C}(\boldsymbol{Z}_{s_0}, \dots, \boldsymbol{Z}_{s_i}, \dots, \boldsymbol{Z}_{s_{n-1}}) = \boldsymbol{0} \end{cases} \tag{30}$$

$$\boldsymbol{C}\left(\boldsymbol{Z}_{S_0}, \dots, \boldsymbol{Z}_{s_i}, \dots, \boldsymbol{Z}_{s_{n-1}}\right) = \begin{bmatrix} \boldsymbol{C}_{s_0}\left(\boldsymbol{Z}_{s_0}, \overline{\boldsymbol{Y}}_{s_0}(\boldsymbol{Z}_{s_0})\right) \\ \boldsymbol{C}_{s_1}\left(\boldsymbol{\Gamma}_{\boldsymbol{Z}_{s_1}}\left(\boldsymbol{\varphi}_{s_1}^{\mathrm{sb}_0}(\boldsymbol{Z}_{s_0})\right), \boldsymbol{\gamma}_{s_1}(\boldsymbol{Z}_{s_1})\right) \\ \vdots \\ \boldsymbol{C}_{s_i}\left(\boldsymbol{\Gamma}_{\boldsymbol{Z}_{s_i}}\left(\boldsymbol{\varphi}_{s_i}^{\mathrm{sb}_{i-1}}(\boldsymbol{Z}_{s_{i-1}})\right), \boldsymbol{\gamma}_{s_i}(\boldsymbol{Z}_{s_i})\right) \\ \vdots \\ \boldsymbol{C}_{s_{n-1}}\left(\boldsymbol{\Gamma}_{\boldsymbol{Z}_{s_{n-1}}}\left(\boldsymbol{\varphi}_{s_{n-1}}^{\mathrm{sb}_{n-2}}(\boldsymbol{Z}_{s_{n-2}})\right), \boldsymbol{\gamma}_{s_{n-1}}(\boldsymbol{Z}_{s_{n-1}})\right) \\ \boldsymbol{C}_{s_n}\left(\boldsymbol{\Gamma}_{\boldsymbol{Y}_{s_n}}\left(\boldsymbol{\varphi}_{s_{n-1}}^{\mathrm{sb}_{n-1}}\left(\boldsymbol{\gamma}_{s_{n-1}}(\boldsymbol{Z}_{s_{n-1}})\right)\right), \overline{\boldsymbol{Y}}_{s_n}\left(\boldsymbol{\Gamma}_{\boldsymbol{X}_{s_n}}\left(\boldsymbol{\varphi}_{s_n}^{\mathrm{sb}_{n-1}}\left(\boldsymbol{\gamma}_{s_{n-1}}(\boldsymbol{Z}_{s_{n-1}})\right)\right)\right)\right) \end{bmatrix} \begin{matrix} \text{start BC} \\ \left.\begin{matrix} \\ \\ \\ \\ \end{matrix}\right\} \text{multishooting} \\ \text{end BC} \end{matrix} \tag{31}$$

All integration domains are independent. Therefore, the integration can be optimized by parallelizing the calculations. The IVP can also be integrated at once by normalizing the curvilinear abscissa such that the integration domain is the $[0,1]$ interval through the change of variable (32). Equation (33) allows us to recalculate the original curvilinear abscissa to evaluate the constitutive laws defined with respect to the zero curvilinear abscissa of the uncut segment.

$$s^i = \frac{s - s_i}{s_{i+1} - s_i} \in [0,1] \tag{32}$$

$$s = s_i + (s_{i+1} - s_i)s^i \tag{33}$$

Note that in (30), in order to have a symmetric constraint vector, we chose to construct $\boldsymbol{\varphi}_{s_0}^{\mathrm{sb}_0} = \boldsymbol{\gamma}_{s_0}\left(\boldsymbol{X}_{s_0}, \boldsymbol{Y}_{s_0}\right)$ without the known part $\overline{\boldsymbol{Y}}_{s_0}$ of the boundary conditions extracted from Table 1 such that $\boldsymbol{Z}_{s_0} = \left[\boldsymbol{X}_{s_0}, \boldsymbol{Y}_{s_0}\right]^{\boldsymbol{T}}$. As described in the single-shooting method section, this is a choice and we could just as easily consider using the shooting method variable $\boldsymbol{Z}_{s_0} = \boldsymbol{X}_{s_0}$, construct $\boldsymbol{\varphi}_{s_0}^{\mathrm{sb}_0} = \boldsymbol{\gamma}_{s_0}\left(\boldsymbol{X}_{s_0}, \overline{\boldsymbol{Y}}_{s_0}\left(\boldsymbol{X}_{s_0}\right)\right)$, and remove the constraint $\boldsymbol{C}_{s_0}$ from $\boldsymbol{C}$.

It is important to note that the segmentation performed in the multi-shooting method is not a discretization method (such as FEM) because all fields remain continuous with respect to the curvilinear abscissa $s$.

## 3.3. Multi-shooting method extended to multi-body

### 3.3.1. Principle

Previous sections dealt with the application of the classical single and multiple shooting methods to the formalism of static slender body TPBVP. The present section discusses the multi-shooting method considering a multi-bodies perspective. This work is motivated by the fact that the classical multi-shooting method does not allow more than two segments to be connected in a simple and generic manner. For example, a three-segment configuration connected to a buoy, such as that described in Figure 5 and discussed in the examples, does not integrate well in the classical multi-shooting method: specific constraint equations must be added to connect the three segments together and apply a vertical force representing the buoy. Therefore, the aim of this section is to propose a generic formalism based on the multi-body philosophy of static bodies connected through kinematic links, while solving the continuous statics of slender bodies through the shooting method.

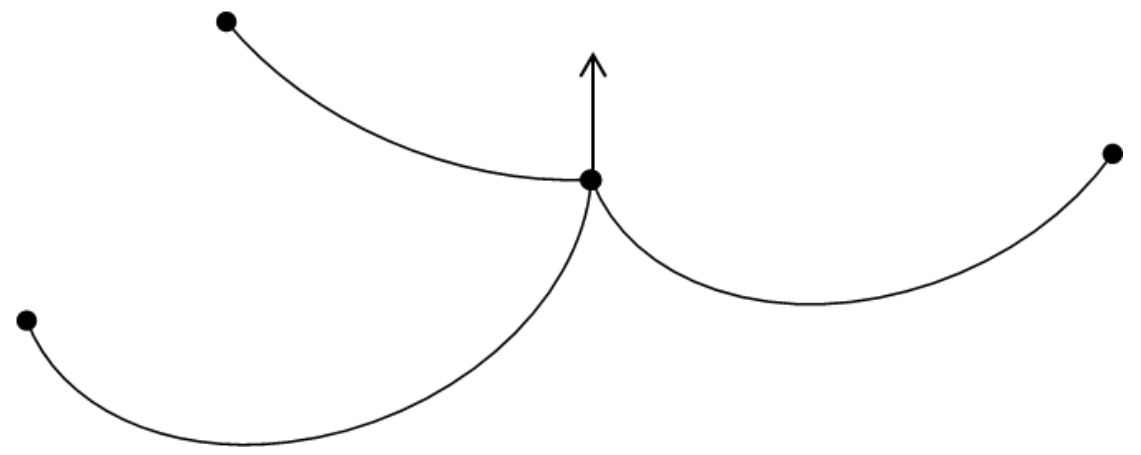

**Figure 5 Example of a three-segment configuration**

Figure 6 shows three configurations: (a) an *open-chain* multi-body configuration, (b) a multi-shooting method, as presented in the previous section, and (c) the proposed mixed multi-body/multi-shooting method.

For the open-chain configuration (a), the rigid body $\mathrm{rb}_j$ is connected to the rigid body $\mathrm{rb}_{j-1}$ by the connection (kinematic link) $\mathrm{c}_{k-1}$, and to the rigid body $\mathrm{rb}_{j+1}$ by the kinematic link $\mathrm{c}_k$. $\boldsymbol{r}_{\mathrm{rb}_j}$ and $R_{\mathrm{rb}_j}$ are respectively the absolute position vector and rotation matrix of the body $\mathrm{rb}_j$. $\boldsymbol{r}_{\mathrm{rb}_j/\mathrm{rb}_{j+1}}$ and $R_{\mathrm{rb}_j/\mathrm{rb}_{j+1}}$ respectively describe the relative position and rotation matrix of the body $\mathrm{rb}_j$ relative to the body $\mathrm{rb}_{j+1}$.

The first idea is to consider a slender body $\mathrm{sb}_i$ as a particular body that can be linked through a kinematic connection, similar to rigid bodies, as shown in diagram (b). The terms $\boldsymbol{r}_{\mathrm{sb}_i/\mathrm{sb}_{i-1}}$, $R_{\mathrm{sb}_i/\mathrm{sb}_{i-1}}$, $\boldsymbol{n}_{\mathrm{sb}_i/\mathrm{sb}_{i-1}}$ and $\boldsymbol{m}_{\mathrm{sb}_i/\mathrm{sb}_{i-1}}$ correspond to the continuous fields $\boldsymbol{r}$, $R$, $\boldsymbol{n}$ and $\boldsymbol{m}$ evaluated at the ends of the slender bodies $\mathrm{sb}_{i-1}$ and $\mathrm{sb}_i$ (i.e., $s = 0$ or $s = l_i$ where $l_i$ is the length of segment $\mathrm{sb}_i$) and differentiated. They correspond to the relative position, orientation, and applied force/moment from one body to the other. Note that in the case of strings kinematics, internal moments $\boldsymbol{m}$ and section orientation $R$ are not among the unknowns of the problem.

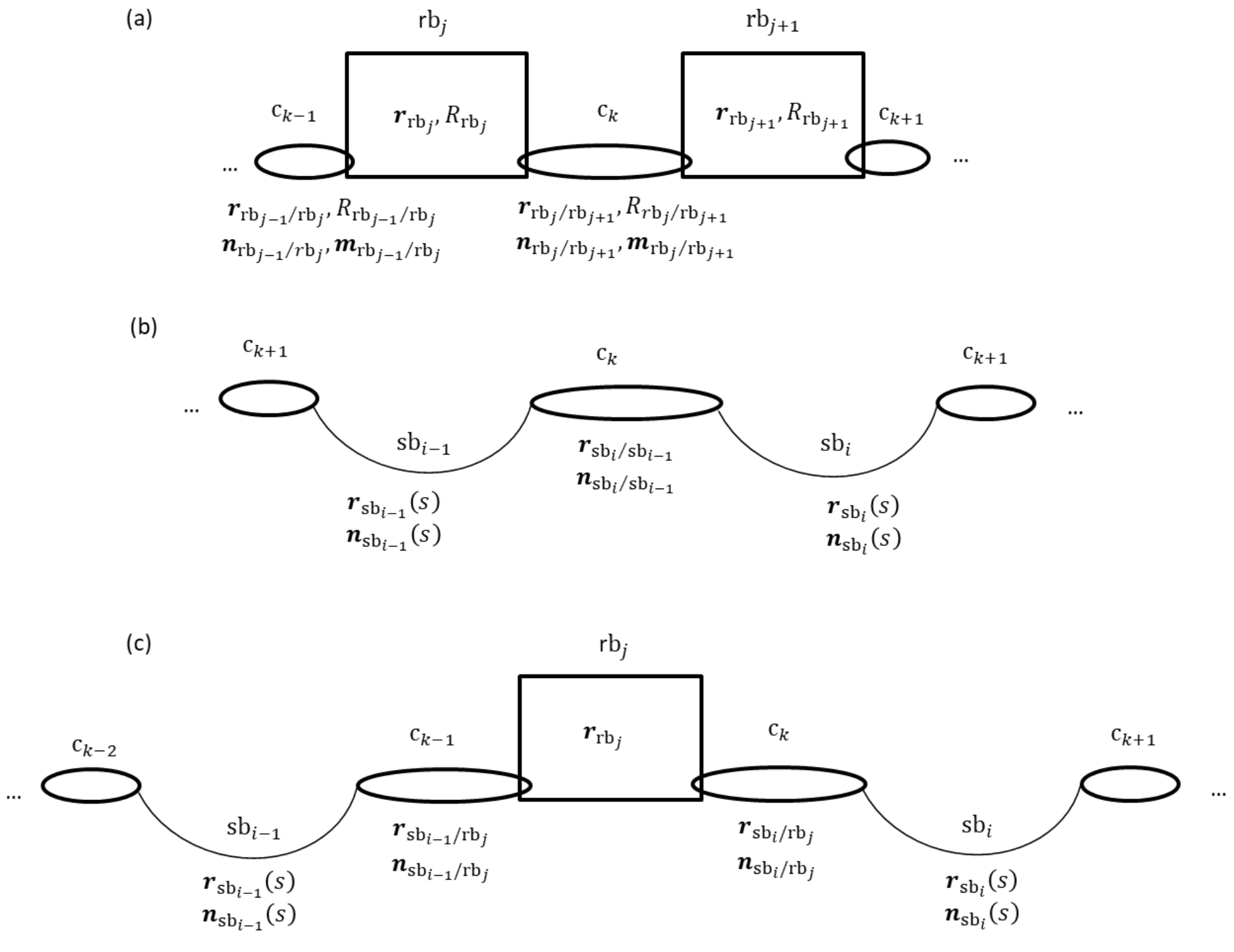

**Figure 6 Open chain multi-body and multiple shooting**

Therefore, the classic multi-shooting method described in the previous section is a special case of configuration (b), where the continuity constraints of the multi-shooting method actually correspond to clamped connections. As previously stated, the disadvantage of configuration (b) is that the segments can only be connected in pairs.

For clarity, the following paragraphs will make use of a local curvilinear abscissa $s \in [0, l_i]$ associated with the segment $\mathrm{sb}_i$. In the case of the multi-shooting method, where a segment is part of a larger slender body, one can retrieve the global curvilinear abscissa by the simple change of variable defined in (33).

Finally, if slender bodies are considered as bodies, they can be connected to rigid bodies, as shown in diagram (c). The connections between the elements (slender and rigid) are the boundary conditions of the slender bodies, and the elements developed in paragraph 2.3 still apply. The structure of (c) allows arbitrary slender body assemblies to be modeled by linking them to a rigid body $\mathrm{rb}_j$ with any type of connection. Figure 7 shows the three-segment configuration depicted in formalism (c). We introduce here the notation to express a connection both attached to a rigid body $\mathrm{rb}_j$ and a slender body $\mathrm{sb}_i$ as $c_k^{\mathrm{rb}_j} = c_{0\,|\,l_i}^{\mathrm{sb}_i}$ where index $k$ refers to the $k\,th$ connection attached to rigid body $\mathrm{rb}_j$, and $0 \mid l_i$ refers to either the connection at $s = 0$ or $s = l_i$ of segment $\mathrm{sb}_i$, with $l_i$ the length of segment $\mathrm{sb}_i$. The relation $c_k^{\mathrm{rb}_j} = c_{0\,|\,l_i}^{\mathrm{sb}_i}$ let us express a mapping between the connections attached to a slender body and a rigid body. To express this mapping, a connection is identified by both reference to $\mathrm{rb}_j$ and $\mathrm{sb}_i$ by notation $c_k^{\mathrm{rb}_j} = c_0^{\mathrm{sb}_i} = c_{k/0}^{\mathrm{rb}_j/\mathrm{sb}_i}$ and $c_k^{\mathrm{rb}_j} = c_{l_i}^{\mathrm{sb}_i} = c_{k/l_i}^{\mathrm{rb}_j/\mathrm{sb}_i}$. To simplify the notations, indexing with the latter form will only be used when indexing may be confusing.

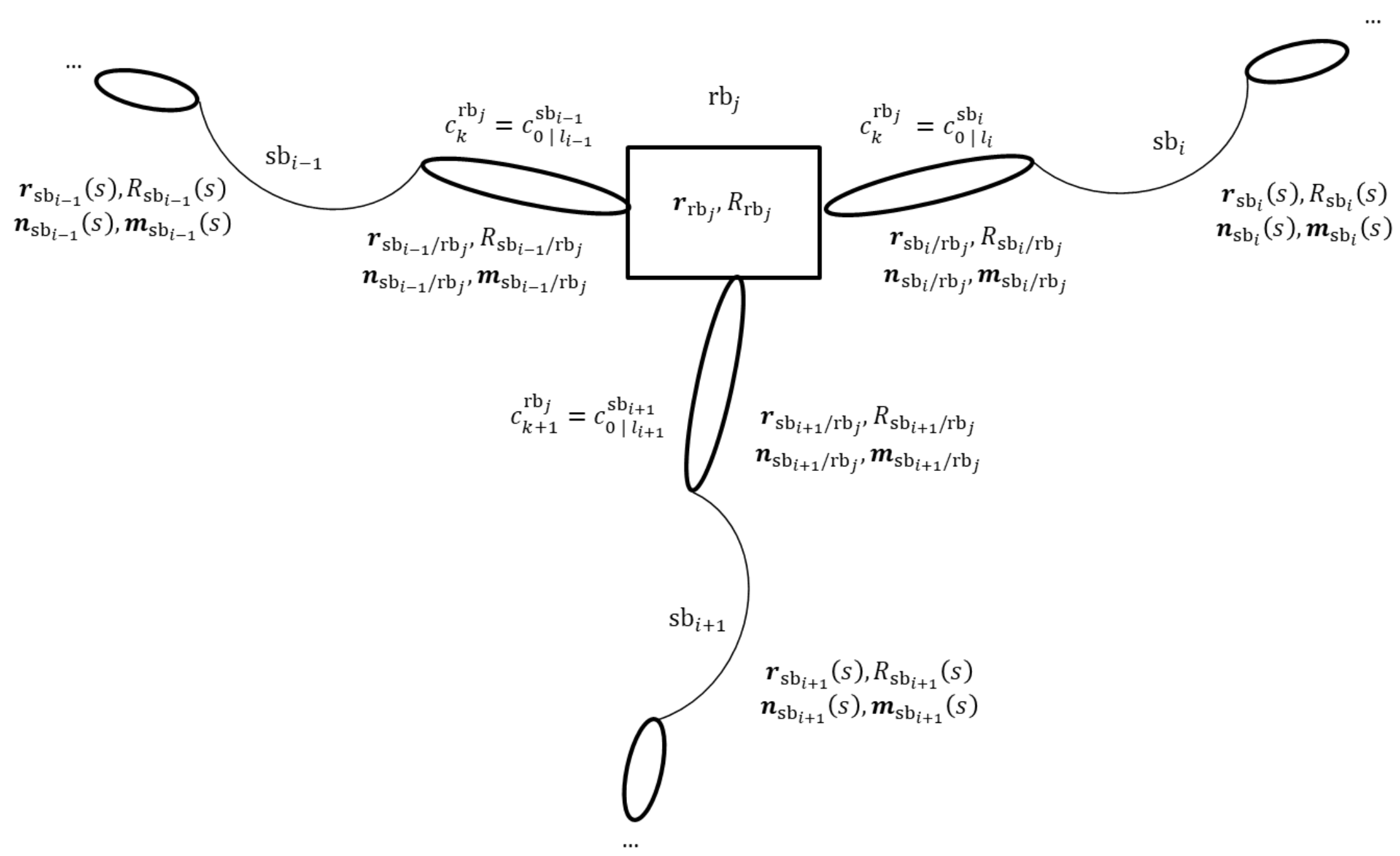

**Figure 7 Example of three segment configuration**

As $c_k^{\mathrm{rb}_j} = c_{0\,|\,l_i}^{\mathrm{sb}_i}$, one can write for strings kinematics the relative position vector as:

$$\boldsymbol{r}_{\mathrm{sb}_i/\mathrm{rb}_j} = \boldsymbol{r}_{0\,|\,l_i}^{\mathrm{sb}_i} - \boldsymbol{r}_{\mathrm{rb}_j} \tag{34}$$

On the other side, the term $\boldsymbol{n}_{\mathrm{sb}_i/\mathrm{rb}_j}$ corresponds to the force exerted by the slender body $\mathrm{sb}_i$ on the rigid body $\mathrm{rb}_j$ at the connection point ($s = 0$ or $s = l_i$).

We made use of the string kinematics once again, allowing us to simplify the formulation because the orientation of the rigid bodies is not among the unknowns of the problem and the internal moments are null.

Now that the main principles of the multi-body/multi-shooting method have been exposed, let's dive in the new set of variables and equations.

### 3.3.2. Equations

#### 3.3.2.1. Variables

In terms of variables, the introduction of rigid bodies kinematics necessitates the introduction of the variables $\boldsymbol{X}_{\mathrm{rb}_j} = \boldsymbol{r}_{\mathrm{rb}_j}$ representing the unknown positions for each rigid body. $\boldsymbol{X}_{\mathrm{rb}_j}$ constitutes a set of additional variables in the zero-search algorithm.

Meanwhile, the terms $\boldsymbol{r}_{\mathrm{sb}_i/\mathrm{rb}_j}$ and $\boldsymbol{n}_{\mathrm{sb}_i/\mathrm{rb}_j}$ previously introduced are variables associated to the link (connection) $c_k^{\mathrm{rb}_j} = c_{0\,|\,l_i}^{\mathrm{sb}_i}$. These variables include both the free and imposed fields components of the connection as described in the boundary conditions section. Note that this is compatible with previous development for single and multiple shooting methods. Indeed, for both these methodologies, $\mathrm{rb}_j$ can actually be considered as a global reference frame to which positions and forces are relative. This leads for the single shooting method: $\left[\boldsymbol{r}_{\mathrm{sb}_i/\mathrm{rb}_j}, \pm\boldsymbol{n}_{\mathrm{sb}_i/\mathrm{rb}_j}\right]^T = \left[\boldsymbol{r}_{\mathrm{sb}_i}, \boldsymbol{n}_{\mathrm{sb}_i}\right]^T = [\boldsymbol{r}(s=0), \boldsymbol{n}(s=0)]^T = [\boldsymbol{r}_0, \boldsymbol{n}_0]^T = \boldsymbol{\varphi}_0 = \boldsymbol{\gamma}_0(\boldsymbol{Z}_0)$ from equation (19). For the multiple shooting method, one can also see that $\left[\boldsymbol{r}_{\mathrm{sb}_i/\mathrm{rb}_j}, \pm\boldsymbol{n}_{\mathrm{sb}_i/\mathrm{rb}_j}\right]^T = [\boldsymbol{r}^i(s=s_i), \boldsymbol{n}^i(s=s_i)]^T = \boldsymbol{\varphi}^{\mathrm{sb}_i}(s=s_i) = \boldsymbol{\varphi}_{s_i}^{\mathrm{sb}_i} = \boldsymbol{\gamma}_{s_i}(\boldsymbol{Z}_{s_i})$ from equation (28). The $\pm$ sign that appears before the forces of the slender body $\mathrm{sb}_i$ acting on the rigid body $\mathrm{rb}_j$ comes from the convention chosen for the curvilinear abscissa and the fact that the internal forces $\boldsymbol{n}^i$ of segment $\mathrm{sb}_i$ represent the forces of the cross section at $s + ds$ on the cross section located at $s$.

Following the formalism introduced previously, the known and unknown parts of $\boldsymbol{r}_{\mathrm{sb}_i/\mathrm{rb}_j}$ and $\boldsymbol{n}_{\mathrm{sb}_i/\mathrm{rb}_j}$ components of a connection $c_k^{\mathrm{rb}_j} = c_{0\,|\,l_i}^{\mathrm{sb}_i} = c_{k/\,0\,|\,l_i}^{\mathrm{rb}_j/\mathrm{sb}_i}$ are written $\boldsymbol{X}_{c_{k/\,0\,|\,l_i}^{\mathrm{rb}_j/\mathrm{sb}_i}}$ and $\boldsymbol{Y}_{c_{k/\,0\,|\,l_i}^{\mathrm{rb}_j/\mathrm{sb}_i}}$ with $\overline{\boldsymbol{Y}}_{c_{k/\,0\,|\,l_i}^{\mathrm{rb}_j/\mathrm{sb}_i}}$the values of $\boldsymbol{Y}_{c_{k/\,0\,|\,l_i}^{\mathrm{rb}_j/\mathrm{sb}_i}}$. With this notation, the term $\boldsymbol{X}_{c_{k/0}^{\mathrm{rb}_j/\mathrm{sb}_i}}$ constitutes the unknowns of the root-finding stage of the shooting method, whereas $\boldsymbol{Y}_{c_{k/l_i}^{\mathrm{rb}_j/\mathrm{sb}_i}}$ and $\overline{\boldsymbol{Y}}_{c_{k/l_i}^{\mathrm{rb}_j/\mathrm{sb}_i}}$ appear in the constraint equations.

The kinematic differential equations being parametrized by the absolute position vector $\boldsymbol{r}$ of a material section, we can make use of (34) to retrieve $\boldsymbol{r}_{0\,|\,l_i}^{\mathrm{sb}_i}$. Then the absolute position of a cross section of a slender body $\mathrm{sb}_i$ attached at $s = 0$ to the rigid body $\mathrm{rb}_j$ is expressed $\boldsymbol{r}_{\mathrm{sb}_i}(s = 0) = \boldsymbol{r}_0^{\mathrm{sb}_i} = \boldsymbol{r}_{\mathrm{sb}_i/\mathrm{rb}_j} + \boldsymbol{r}_{\mathrm{rb}_j}$. This relationship makes the state vector $\boldsymbol{\varphi}$ dependent on $\boldsymbol{X}_{rb_j}$ through mappings $\boldsymbol{\gamma}$ and $\boldsymbol{\Gamma}$ such that:

$$\boldsymbol{\varphi}_{c_{0\,|\,l_i}^{\mathrm{sb}_i}} = \boldsymbol{\gamma}_{c_{k/\,0\,|\,l_i}^{\mathrm{rb}_j/\mathrm{sb}_i}}\left(\boldsymbol{X}_{c_{k/\,0\,|\,l_i}^{\mathrm{rb}_j/\mathrm{sb}_i}}, \boldsymbol{Y}_{c_{k/\,0\,|\,l_i}^{\mathrm{rb}_j/\mathrm{sb}_i}}, \boldsymbol{X}_{\mathrm{rb}_j}\right) \tag{35}$$

$$\boldsymbol{X}_{c_{k/\,0\,|\,l_i}^{\mathrm{rb}_j/\mathrm{sb}_i}} = \boldsymbol{\Gamma}_{\boldsymbol{X}_{c_{k/\,0\,|\,l_i}^{\mathrm{rb}_j/\mathrm{sb}_i}}}\left(\boldsymbol{\varphi}_{c_{0\,|\,l_i}^{\mathrm{sb}_i}}, \boldsymbol{X}_{\mathrm{rb}_j}\right) \tag{36}$$

$$\boldsymbol{Y}_{c_{k/\,0\,|\,l_i}^{\mathrm{rb}_j/\mathrm{sb}_i}} = \boldsymbol{\Gamma}_{\boldsymbol{Y}_{c_{k/\,0\,|\,l_i}^{\mathrm{rb}_j/\mathrm{sb}_i}}}\left(\boldsymbol{\varphi}_{c_{0\,|\,l_i}^{\mathrm{sb}_i}}, \boldsymbol{X}_{\mathrm{rb}_j}\right) \tag{37}$$

To consider the case in which the imposed field values depend on unknown fields, the imposed field values are written as:

$$\overline{\boldsymbol{Y}}_{c_{k/\,0\,|\,l_i}^{\mathrm{rb}_j/\mathrm{sb}_i}} = \overline{\boldsymbol{Y}}_{c_{k/\,0\,|\,l_i}^{\mathrm{rb}_j/\mathrm{sb}_i}}\left(\boldsymbol{X}_{c_{k/\,0\,|\,l_i}^{\mathrm{rb}_j/\mathrm{sb}_i}}, \boldsymbol{X}_{\mathrm{rb}_j}\right) \tag{38}$$

Figure 8 summarizes the variables of the multi-body/multi-shooting method for a unitary segment system:

The variables associated with the root finding algorithm are:

- $\boldsymbol{X}_{\mathrm{rb}_j}$ and $\boldsymbol{X}_{\mathrm{rb}_j}$ describing the positions of the rigid bodies respectively at start and end of the segment,

- $\boldsymbol{Z}_{c_{k/0}^{\mathrm{rb}_j/\mathrm{sb}_i}}$ with $\boldsymbol{Z}_{c_{k/0}^{\mathrm{rb}_j/\mathrm{sb}_i}} = \boldsymbol{X}_{c_{k/0}^{\mathrm{rb}_j/\mathrm{sb}_i}}$, or variable $\boldsymbol{Z}_{c_{k/0}^{\mathrm{rb}_j/\mathrm{sb}_i}} = \left[\boldsymbol{X}_{c_{k/0}^{\mathrm{rb}_j/\mathrm{sb}_i}}, \boldsymbol{Y}_{c_{k/0}^{\mathrm{rb}_j/\mathrm{sb}_i}}\right]^T$ if associated with constraint $\boldsymbol{C}_{c_{k/0}^{\mathrm{rb}_j/\mathrm{sb}_i}}\left(\boldsymbol{Z}_{c_{k/0}^{\mathrm{rb}_j/\mathrm{sb}_i}}, \overline{\boldsymbol{Y}}_{c_{k/0}^{\mathrm{rb}_j/\mathrm{sb}_i}}\left(\boldsymbol{Z}_{c_{k/0}^{\mathrm{rb}_j/\mathrm{sb}_i}}, \boldsymbol{X}_{\mathrm{rb}_j}\right)\right)$, which describes the free components of a boundary condition.

The variables associated with the IVP are:

- the state vector field of the segment $\mathrm{sb}_i$ is $\boldsymbol{\varphi}_{\mathrm{sb}_i}(s)$ estimated by the IVP (24),
- whose initial evaluation $\boldsymbol{\varphi}_{c_0^{\mathrm{sb}_i}}$ at $s = 0$ for connection $c_0^{\mathrm{sb}_i} = c_k^{\mathrm{rb}_j} = c_{k/0}^{\mathrm{rb}_j/\mathrm{sb}_i}$ depends on the variables of the root finding stage $\boldsymbol{X}_{\mathrm{rb}_j}$ and $\boldsymbol{Z}_{c_{k/0}^{\mathrm{rb}_j/\mathrm{sb}_i}}$ thanks to mapping (35),
- and whose final evaluation $\boldsymbol{\varphi}_{c_{l_i}^{\mathrm{sb}_i}}$ depends on the state vector $\boldsymbol{\varphi}_{c_0^{\mathrm{sb}_i}}$ of the initial connection $c_0^{\mathrm{sb}_i}$ integrated thanks to (24).

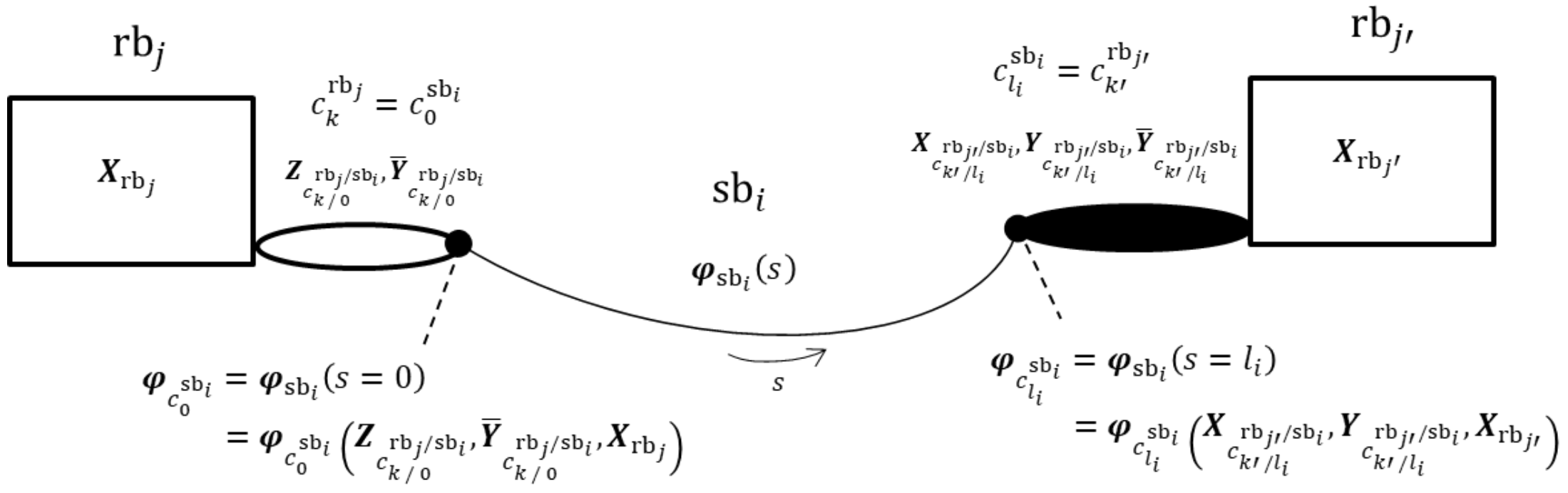

**Figure 8 Unitary kinematic chain**

#### 3.3.2.2. Constraints

Now that we have clarified the expression of both the root finding and IVP problem variables, let us turn to the formulation of the constraint equations of the root finding stage. The constraints come from two sources: as in single shooting method, from the ending boundary conditions constraints $c_{l_i}^{\mathrm{sb}_i} = c_{k'}^{\mathrm{rb}_{j'}}$ (and eventually $c_0^{\mathrm{sb}_i} = c_k^{\mathrm{rb}_j}$ if $\boldsymbol{Z}_{c_{k/0}^{\mathrm{rb}_j/\mathrm{sb}_i}} = \left[\boldsymbol{X}_{c_{k/0}^{\mathrm{rb}_j/\mathrm{sb}_i}}, \boldsymbol{Y}_{c_{k/0}^{\mathrm{rb}_j/\mathrm{sb}_i}}\right]^T$ is chosen), but also from the rigid bodies themselves on which are connected several kinematic links.

The constraint equations coming from boundary conditions attached to segment $\mathrm{sb}_i$ are expressed in the same form as for the simple shooting method. Depending on the choice of unknowns $\boldsymbol{Z}_{c_{k/0}^{\mathrm{rb}_j/\mathrm{sb}_i}} = \boldsymbol{X}_{c_{k/0}^{\mathrm{rb}_j/\mathrm{sb}_i}}$ or $\boldsymbol{Z}_{c_{k/0}^{\mathrm{rb}_j/\mathrm{sb}_i}} = \left[\boldsymbol{X}_{c_{k/0}^{\mathrm{rb}_j/\mathrm{sb}_i}}, \boldsymbol{Y}_{c_{k/0}^{\mathrm{rb}_j/\mathrm{sb}_i}}\right]^T$, the constraint equations from equations (25) and (26) at connections $c_{k/0}^{\mathrm{rb}_j/\mathrm{sb}_i}$ and $c_{k'/l_i}^{\mathrm{rb}_{j'}/\mathrm{sb}_i}$ are either:

$$\boldsymbol{C}_{c_{k'/l_i}^{\mathrm{rb}_{j'}/\mathrm{sb}_i}}\left(\boldsymbol{Z}_{c_{k/0}^{\mathrm{rb}_j/\mathrm{sb}_i}}, \boldsymbol{X}_{\mathrm{rb}_j}, \boldsymbol{X}_{\mathrm{rb}_{j'}}\right) =$$

$$\boldsymbol{C}_{c_{k'/l_i}^{\mathrm{rb}_{j'}/\mathrm{sb}_i}}\begin{pmatrix} \boldsymbol{Y}_{c_{k'/l_i}^{\mathrm{rb}_{j'}/\mathrm{sb}_i}}\left(\boldsymbol{\varphi}_{c_{l_i}^{\mathrm{sb}_i}}\left(\boldsymbol{\varphi}_{c_0^{\mathrm{sb}_i}}\left(\boldsymbol{Z}_{c_{k/0}^{\mathrm{rb}_j/\mathrm{sb}_i}}, \overline{\boldsymbol{Y}}_{c_{k/0}^{\mathrm{rb}_j/\mathrm{sb}_i}}\left(\boldsymbol{Z}_{c_{k/0}^{\mathrm{rb}_j/\mathrm{sb}_i}}, \boldsymbol{X}_{\mathrm{rb}_j}\right), \boldsymbol{X}_{\mathrm{rb}_j}\right)\right), \boldsymbol{X}_{\mathrm{rb}_{j'}}\right), \\ \overline{\boldsymbol{Y}}_{c_{k'/l_i}^{\mathrm{rb}_{j'}/\mathrm{sb}_i}}\left(\boldsymbol{X}_{c_{k'/l_i}^{\mathrm{rb}_{j'}/\mathrm{sb}_i}}\left(\boldsymbol{\varphi}_{c_{l_i}^{\mathrm{sb}_i}}\left(\boldsymbol{\varphi}_{c_0^{\mathrm{sb}_i}}\left(\boldsymbol{Z}_{c_{k/0}^{\mathrm{rb}_j/\mathrm{sb}_i}}, \overline{\boldsymbol{Y}}_{c_{k/0}^{\mathrm{rb}_j/\mathrm{sb}_i}}\left(\boldsymbol{Z}_{c_{k/0}^{\mathrm{rb}_j/\mathrm{sb}_i}}, \boldsymbol{X}_{\mathrm{rb}_j}\right), \boldsymbol{X}_{\mathrm{rb}_j}\right)\right), \boldsymbol{X}_{\mathrm{rb}_{j'}}\right), \boldsymbol{X}_{\mathrm{rb}_{j'}}\right), \\ \boldsymbol{X}_{\mathrm{rb}_{j'}} \end{pmatrix} \tag{39}$$

Or:

$$\boldsymbol{C}_{\mathrm{sb}_i}\left(\boldsymbol{Z}_{c_{k/0}^{\mathrm{rb}_j/\mathrm{sb}_i}}, \boldsymbol{X}_{\mathrm{rb}_j}, \boldsymbol{X}_{\mathrm{rb}_{j'}}\right) =$$

$$\begin{cases} \boldsymbol{C}_{c_{k/0}^{\mathrm{rb}_j/\mathrm{sb}_i}}\left(\boldsymbol{Z}_{c_{k/0}^{\mathrm{rb}_j/\mathrm{sb}_i}}, \overline{\boldsymbol{Y}}_{c_{k/0}^{\mathrm{rb}_j/\mathrm{sb}_i}}\left(\boldsymbol{Z}_{c_{k/0}^{\mathrm{rb}_j/\mathrm{sb}_i}}, \boldsymbol{X}_{\mathrm{rb}_j}\right), \boldsymbol{X}_{\mathrm{rb}_j}\right) \\ \boldsymbol{C}_{c_{k'/l_i}^{\mathrm{rb}_{j'}/\mathrm{sb}_i}}\begin{pmatrix} \boldsymbol{Y}_{c_{k'/l_i}^{\mathrm{rb}_{j'}/\mathrm{sb}_i}}\left(\boldsymbol{\varphi}_{c_{l_i}^{\mathrm{sb}_i}}\left(\boldsymbol{\varphi}_{c_0^{\mathrm{sb}_i}}\left(\boldsymbol{Z}_{c_{k/0}^{\mathrm{rb}_j/\mathrm{sb}_i}}, \boldsymbol{X}_{\mathrm{rb}_j}\right)\right), \boldsymbol{X}_{\mathrm{rb}_{j'}}\right), \\ \overline{\boldsymbol{Y}}_{c_{k'/l_i}^{\mathrm{rb}_{j'}/\mathrm{sb}_i}}\left(\boldsymbol{X}_{c_{k'/l_i}^{\mathrm{rb}_{j'}/\mathrm{sb}_i}}\left(\boldsymbol{\varphi}_{c_{l_i}^{\mathrm{sb}_i}}\left(\boldsymbol{\varphi}_{c_0^{\mathrm{sb}_i}}\left(\boldsymbol{Z}_{c_{k/0}^{\mathrm{rb}_j/\mathrm{sb}_i}}, \boldsymbol{X}_{\mathrm{rb}_j}\right)\right), \boldsymbol{X}_{\mathrm{rb}_{j'}}\right), \boldsymbol{X}_{\mathrm{rb}_{j'}}\right), \\ \boldsymbol{X}_{\mathrm{rb}_{j'}} \end{pmatrix} \end{cases} \tag{40}$$

In both cases $\boldsymbol{C}_{c_{k'/l_i}^{\mathrm{rb}_{j'}/\mathrm{sb}_i}} = \boldsymbol{C}_{c_{k'/l_i}^{\mathrm{rb}_{j'}/\mathrm{sb}_i}}\left(\boldsymbol{Z}_{c_{k/0}^{\mathrm{rb}_j/\mathrm{sb}_i}}, \boldsymbol{X}_{\mathrm{rb}_j}, \boldsymbol{X}_{\mathrm{rb}_{j'}}\right)$ while $\boldsymbol{C}_{c_{k/0}^{\mathrm{rb}_j/\mathrm{sb}_i}} = \boldsymbol{C}_{c_{k/0}^{\mathrm{rb}_j/\mathrm{sb}_i}}\left(\boldsymbol{Z}_{c_{k/0}^{\mathrm{rb}_j/\mathrm{sb}_i}}, \boldsymbol{X}_{\mathrm{rb}_j}\right)$. This particular structure allows an immediate comparison with the mathematical developments made in the section on the single-shooting method and preserves the associated interpretations and conclusions.

The static equilibrium of all forces transmitted by the $p$ connections $c_k^{\mathrm{rb}_j} = c_{0\,|\,l_i}^{\mathrm{sb}_i}$ attached to the rigid body $\mathrm{rb}_j$ yield:

$$\sum_i \boldsymbol{n}_{\mathrm{sb}_i/\mathrm{rb}_j} - F\left(\boldsymbol{r}_{\mathrm{rb}_j}\right) = 0 \tag{41}$$

Here, we define a mapping from the IVP problem to the root finding stage for this latest constraint. Let us introduce the term $\boldsymbol{Y}_{c_k^{\mathrm{rb}_j}}$ as the extraction of constrained fields $\boldsymbol{n}_{\mathrm{sb}_i/\mathrm{rb}_j}$ from the state vector $\boldsymbol{\varphi}_{c_{k\,/\,0\,|\,l_i}^{\mathrm{rb}_j/\mathrm{sb}_i}}$ at link $c_k^{\mathrm{rb}_j} = c_{0\,|\,l_i}^{\mathrm{sb}_i}$. $\boldsymbol{Y}_{c_k^{\mathrm{rb}_j}}$ is computed thanks to the mapping $\boldsymbol{\Gamma}_{\boldsymbol{Y}_{\mathrm{rb}_j}}$ in equation (42). Note that $\boldsymbol{Y}_{c_k^{\mathrm{rb}_j}} \neq \boldsymbol{Y}_{c_{k/\,0\,|\,l_i}^{\mathrm{rb}_j/\mathrm{sb}_i}}$ because mappings $\boldsymbol{\Gamma}_{\boldsymbol{Y}_{\mathrm{rb}_j}}$ and $\boldsymbol{\Gamma}_{\boldsymbol{Y}_{c_{k/\,0\,|\,l_i}^{\mathrm{rb}_j/\mathrm{sb}_i}}}$ are note equal (i.e., do not extract the same parts of $\boldsymbol{\varphi}$), even if the connection is identified as either attached to a rigid body or a slender body by its index $c_k^{\mathrm{rb}_j} = c_{0\,|\,l_i}^{\mathrm{sb}_i}$.

The term $\overline{\boldsymbol{Y}}_{\mathrm{rb}_j}$ expressed in (43) for a rigid body is equivalent to (38) for a connection. It represents the value of forces $F\left(\boldsymbol{r}_{\mathrm{rb}_j}\right)$ other than those arising from connections, which may depend on the rigid body position.

$$\boldsymbol{Y}_{c_k^{\mathrm{rb}_j}} = \boldsymbol{\Gamma}_{\boldsymbol{Y}_{\mathrm{rb}_j}}\left(\boldsymbol{\varphi}_{c_{k/0\,|\,l_i}^{\mathrm{rb}_j/\mathrm{sb}_i}}, \boldsymbol{X}_{\mathrm{rb}_j}\right) \tag{42}$$

$$\overline{\boldsymbol{Y}}_{\mathrm{rb}_j} = \overline{\boldsymbol{Y}}_{\mathrm{rb}_j}\left(\boldsymbol{X}_{\mathrm{rb}_j}\right) \tag{43}$$

These notations allow us to rewrite the static equilibrium constraint of rigid body $\mathrm{rb}_j$ (41) in terms of the root finding variables:

$$\boldsymbol{C}_{\mathrm{rb}_j}\left(\sum_k \boldsymbol{Y}_{c_k^{\mathrm{rb}_j}}, \overline{\boldsymbol{Y}}_{\mathrm{rb}_j}\left(\boldsymbol{X}_{\mathrm{rb}_j}\right)\right) = \sum_k \boldsymbol{Y}_{c_k^{\mathrm{rb}_j}} - \overline{\boldsymbol{Y}}_{\mathrm{rb}_j}\left(\boldsymbol{X}_{\mathrm{rb}_j}\right) = 0 \tag{44}$$

The term $\boldsymbol{Y}_{c_k^{\mathrm{rb}_j}}$ shall be expressed with regard to the root finding variables $\boldsymbol{X}_{\mathrm{rb}_j}$ and $\boldsymbol{Z}_{c_{k/0}^{\mathrm{rb}_j/\mathrm{sb}_i}}$:

$$\begin{aligned}\sum_k \boldsymbol{Y}_{c_k^{\mathrm{rb}_j}} &= \sum_k \boldsymbol{Y}_{c_{k,\alpha}^{\mathrm{rb}_j}} + \sum_k \boldsymbol{Y}_{c_{k,\omega}^{\mathrm{rb}_j}} = \boldsymbol{Y}_{\Sigma c_\alpha^{\mathrm{rb}_j}} + \boldsymbol{Y}_{\Sigma c_\omega^{\mathrm{rb}_j}} = \\ &= \sum_k \boldsymbol{\Gamma}_{\boldsymbol{Y}_{\mathrm{rb}_j}}\left(\boldsymbol{\varphi}_{c_{k/0}^{\mathrm{rb}_j/\mathrm{sb}_i}}\left(\boldsymbol{Z}_{c_{k/0}^{\mathrm{rb}_j/\mathrm{sb}_i}}, \boldsymbol{X}_{\mathrm{rb}_j}\right), \boldsymbol{X}_{\mathrm{rb}_j}\right) \\ &+ \sum_k \boldsymbol{\Gamma}_{\boldsymbol{Y}_{\mathrm{rb}_j}}\left(\boldsymbol{\varphi}_{c_{k/l_{i'}}^{\mathrm{rb}_j/\mathrm{sb}_{i'}}}\left(\boldsymbol{\varphi}_{c_0^{\mathrm{sb}_{i'}}}\left(\boldsymbol{Z}_{c_{k'/0}^{\mathrm{rb}_{j'}/\mathrm{sb}_{i'}}}, \boldsymbol{X}_{\mathrm{rb}_{j'}}\right)\right), \boldsymbol{X}_{\mathrm{rb}_j}\right)\end{aligned} \tag{45}$$

Where we splitted the terms coming from connections that start a slender body $c_{k,\alpha}^{\mathrm{rb}_j}$ with connections that end a slender body $c_{k,\omega}^{\mathrm{rb}_j}$. We chose to remove the explicit reference of notation $c_{k/0\,|\,l_i}^{\mathrm{rb}_j/\mathrm{sb}_i}$ to the segment to prefer the lighter notation $c_{k,\alpha}^{\mathrm{rb}_j}$ or $c_{k,\omega}^{\mathrm{rb}_j}$. We made this choice to avoid confusion on segment indexing in the sums appearing in (45), because each connection $c_k^{\mathrm{rb}_j}$ , where $c_k^{\mathrm{rb}_j} = c_{0\,|\,l_i}^{\mathrm{sb}_i}$, is attached to a different slender body $\mathrm{sb}_i$. Thus, the sum on index $k$ is equivalent to a sum on index $i$ limited to segments attached to the rigid body $\mathrm{rb}_j$. Then, notation $c_{k,\alpha}^{\mathrm{rb}_j}$ designates all segment connections that satisfy $c_k^{\mathrm{rb}_j} = c_0^{\mathrm{sb}_i}$, whereas notation $c_{k,\omega}^{\mathrm{rb}_j}$ designates all segment connections that satisfy $c_k^{\mathrm{rb}_j} = c_{l_i}^{\mathrm{sb}_i}$.

Note that the mapping $\boldsymbol{\Gamma}_{\boldsymbol{Y}_{\mathrm{rb}_j}}$ is the same for extracting $\boldsymbol{Y}_{c_{k,\alpha}^{\mathrm{rb}_j}}$ and $\boldsymbol{Y}_{c_{k,\omega}^{\mathrm{rb}_j}}$. The only difference is its application to either $\boldsymbol{\varphi}_{c_0^{\mathrm{sb}_i}}$ or $\boldsymbol{\varphi}_{c_{l_{i'}}^{\mathrm{sb}_{i'}}}$. It follows that:

$$\boldsymbol{C}_{\mathrm{rb}_j} = \boldsymbol{C}_{\mathrm{rb}_j}\left(\boldsymbol{Z}_{c_{k/0}^{\mathrm{rb}_j/\mathrm{sb}_i}}, \boldsymbol{X}_{\mathrm{rb}_j}, \boldsymbol{Z}_{c_{k'/0}^{\mathrm{rb}_{j'}/\mathrm{sb}_{i'}}}, \boldsymbol{X}_{\mathrm{rb}_{j'}}\right)$$

$$= \boldsymbol{C}_{\mathrm{rb}_j}\left(\sum_k \boldsymbol{Y}_{c_{k,\alpha}^{\mathrm{rb}_j}}\left(\boldsymbol{\varphi}_{c_0^{\mathrm{sb}_i}}\left(\boldsymbol{Z}_{c_{k/0}^{\mathrm{rb}_j/\mathrm{sb}_i}}, \boldsymbol{X}_{\mathrm{rb}_j}\right), \boldsymbol{X}_{\mathrm{rb}_j}\right), \sum_k \boldsymbol{Y}_{c_{k,\omega}^{\mathrm{rb}_j}}\left(\boldsymbol{\varphi}_{c_{l_{i'}}^{\mathrm{sb}_{i'}}}\left(\boldsymbol{\varphi}_{c_0^{\mathrm{sb}_{i'}}}\left(\boldsymbol{Z}_{c_{k'/0}^{\mathrm{rb}_{j'}/\mathrm{sb}_{i'}}}, \boldsymbol{X}_{\mathrm{rb}_{j'}}\right)\right), \boldsymbol{X}_{\mathrm{rb}_j}\right), \bar{\boldsymbol{Y}}_{\mathrm{rb}_j}\left(\boldsymbol{X}_{\mathrm{rb}_j}\right)\right)$$

It should be noted that contrary to the constraint equations generated by the boundary conditions $\boldsymbol{C}_{c_{k'/l_i}^{\mathrm{rb}_{j'}/\mathrm{sb}_i}}\left(\boldsymbol{Z}_{c_{k/0}^{\mathrm{rb}_j/\mathrm{sb}_i}}, \boldsymbol{X}_{\mathrm{rb}_j}, \boldsymbol{X}_{\mathrm{rb}_{j'}}\right)$ (39) and $\boldsymbol{C}_{c_{k/0}^{\mathrm{rb}_j/\mathrm{sb}_i}}\left(\boldsymbol{Z}_{c_{k/0}^{\mathrm{rb}_j/\mathrm{sb}_i}}, \boldsymbol{X}_{\mathrm{rb}_j}\right)$ (40) which where depending on the unknowns of a single connection, $\boldsymbol{C}_{\mathrm{rb}_j}$ introduces couplings between the unknowns of several connections attached to different segments, all linked together by the same rigid body $\mathrm{rb}_j$.

In formalism (c), the rigid body position variables are considered as new unknowns, and equilibrium equations (44) constitute new constraints that are concatenated to the constraint vector $\boldsymbol{C}$.

We can interpret the proposed formalism (c) within the framework of the classical multi-shooting method formalism (b), in which all positions and internal forces fields are constrained with a unique continuity equation at each segment junction. In formalism (c), this unique constraint is split into three parts: two clamped connections (which impose the positions of the segment's ends with respect to the rigid body's position, and leave unconstrained the internal forces), with a rigid body static equilibrium that equilibrates the internal forces from both segments.

Compared to formalism (b), cutting a segment into two sub-segments with formalism (c) also generates 6 unknowns and 6 new constraints. The size of the problem is kept identical, while the proposed formalism (c) allows dealing with any type of kinematic link between slender and rigid bodies, as well as connecting several slender bodies to a single rigid body.

The general formulation of the mixed multi-body/multi-shooting constraints for $n$ segments ($2n$ connections) and $m$ bodies is given by equation (46):

$$\begin{cases} \begin{cases} \partial_s \boldsymbol{\phi}(s) = \boldsymbol{G}(\boldsymbol{\phi}, s) \\ \boldsymbol{\phi}_0 = \left[\cdots, \quad \boldsymbol{\gamma}_{c_{k/0}^{\mathrm{rb}_j/\mathrm{sb}_i}}\left(\boldsymbol{Z}_{c_{k/0}^{\mathrm{rb}_j/\mathrm{sb}_i}}, \boldsymbol{X}_{\mathrm{rb}_j}\right), \quad \cdots\right]^T \end{cases} \\ \boldsymbol{C}\left(\dots, \boldsymbol{Z}_{c_{k/0}^{\mathrm{rb}_j/\mathrm{sb}_i}}, \dots, \boldsymbol{X}_{\mathrm{rb}_j}, \dots\right) = \boldsymbol{0} \end{cases} \tag{46}$$

Where $\boldsymbol{\phi}(s) = \left[\boldsymbol{\varphi}_{\mathrm{sb}_0}(s), \cdots, \boldsymbol{\varphi}_{\mathrm{sb}_i}(s), \cdots, \boldsymbol{\varphi}_{\mathrm{sb}_{n-1}}(s)\right]^T$ and:

$$\boldsymbol{G}(\boldsymbol{\phi}, s) = \begin{bmatrix} \boldsymbol{F}\left(\boldsymbol{\varphi}_{\mathrm{sb}_0}, s\right) \\ \vdots \\ \boldsymbol{F}\left(\boldsymbol{\varphi}_{\mathrm{sb}_i}, s\right) \\ \vdots \\ \boldsymbol{F}\left(\boldsymbol{\varphi}_{\mathrm{sb}_{n-1}}, s\right) \end{bmatrix}$$

$$\boldsymbol{C}\left(\dots, \boldsymbol{Z}_{c_{k/0}^{\mathrm{rb}_j/\mathrm{sb}_i}}, \dots, \boldsymbol{X}_{\mathrm{rb}_j}, \dots\right) = \begin{bmatrix} \boldsymbol{C}_{c_{k/0}^{\mathrm{rb}_j/\mathrm{sb}_i}}\left(\boldsymbol{Z}_{c_{k/0}^{\mathrm{rb}_j/\mathrm{sb}_i}}, \boldsymbol{X}_{\mathrm{rb}_j}\right) \\ \boldsymbol{C}_{c_{k'/l_i}^{\mathrm{rb}_{j'}/\mathrm{sb}_i}}\left(\boldsymbol{Z}_{c_{k/0}^{\mathrm{rb}_j/\mathrm{sb}_i}}, \boldsymbol{X}_{\mathrm{rb}_j}, \boldsymbol{X}_{\mathrm{rb}_{j'}}\right) \\ \boldsymbol{C}_{\mathrm{rb}_j}\left(\boldsymbol{Z}_{c_{k/0}^{\mathrm{rb}_j/\mathrm{sb}_i}}, \boldsymbol{X}_{\mathrm{rb}_j}, \boldsymbol{Z}_{c_{k'/0}^{\mathrm{rb}_{j'}/\mathrm{sb}_{i'}}}, \boldsymbol{X}_{\mathrm{rb}_{j'}}\right) \end{bmatrix}$$

With:

- $\boldsymbol{C}_{c_{k/0}^{\mathrm{rb}_j/\mathrm{sb}_i}}$ the constraint generated by the connection $c_{k/0}^{\mathrm{rb}_j/\mathrm{sb}_i}$ which starts the segment $\mathrm{sb}_i$, also attached to the rigid body $\mathrm{rb}_j$,
- $\boldsymbol{C}_{c_{k/l_i}^{\mathrm{rb}_j/\mathrm{sb}_i}}$ the constraint generated by connection $c_{k/l_i}^{\mathrm{rb}_j/\mathrm{sb}_i}$ which ends the segment $\mathrm{sb}_i$, also attached to the rigid body $\mathrm{rb}_j$,
- $\boldsymbol{C}_{\mathrm{rb}_j}$ the constraints generated by the rigid body $\mathrm{rb}_j$.

One may keep in mind that indices $j$ and $j'$, and $i$ and $i'$ are dummy indices in the expressions of $\boldsymbol{C}$, and solely express the distinction between several rigid/slender bodies.

## 4. Numerical experimentations

This section presents the results obtained from the numerical experiments of the multi-body/multi-shooting methods developed in the previous section for string kinematics. The objectives are multiple: to demonstrate the capabilities of the method applied to string problems in terms of:

- Precision measured by comparing the results to a reference (analytical, or semi- analytical),
- Convergence measured by the number of Newton iterations and number of nodes in adaptive step integration,
- Modeling capabilities for strings: single or multi-segment, different boundary conditions, different material section parameters or different types of constitutive laws, and nonlinear external linear loading (depending on the geometry of the solution).

References for the comparison of results are taken either from analytical (or semi-analytical: numerically solved) formulations or from the literature. The four examples summarized in Table 2 were treated to cover all these topics. The types of references considered are also listed in this table.

| | 1 | 2 | 3 | 4 |
|---|---|---|---|---|
| Reference | C | C | A | C |
| Single segment | X | X | X | |
| Multi segments | | | | X |
| Boundary conditions | X | X | | |
| Constitutive law | | | X | X |
| Convergence | X | X | | |
| Integrator (number of nodes) | X | | | X |
| Non-linear loading (multiple solutions) | | | X | |
| A: analytical<br>C: catenary (semi-analytical) | | | | |

**Table 2: Shooting method - validation examples**

Two references are used in this study:

- Catenary formulation [21] This is actually a derivative of the shooting method, for which the ODE has been integrated analytically (see Appendix: semi-analytical reference for catenary string examples). Once integrated, the equations form an algebraic system to be solved. As with the shooting method, a Newton algorithm is used,
- Analytical formulation of the nonlinear *Velaria* problem [16].

### 4.1. Example 1: catenary string with ball-prismatic joints

This example studies a single string subject only to its own weight (catenary) attached at the ends by a ball-and-socket joint and a prismatic joint of axis $\boldsymbol{x}$ (Figure 9). This problem enabled us to verify the shooting method, in which several values of the horizontal force of the prismatic connection were targeted. The aim is to break out of the conventional catenary modeling framework in which the segment is connected by ball-and-socket joints at each end. In this validation example, we also examine the influence of the Runge-Kutta integration scheme error on the number of points and accuracy.

The segment was assumed to have a uniform material cross-section, whose parameters are listed in Table 3. The line is fixed in $s = 0$ and is free to translate along $\boldsymbol{x}$ at $s = L$ with an imposed force $F_x$.

$V_a$, $H_a$, $V$ and $H$ are respectively the vertical and horizontal tensions at the left (anchor) and right (fairlead) ends. Newton's tolerance is set at $10^{-12}$ and initially, we use an explicit RK4 scheme where we set the number of elements to 10.

The results were compared with the analytical formulation developed in Appendix: semi-analytical reference for catenary string examples.

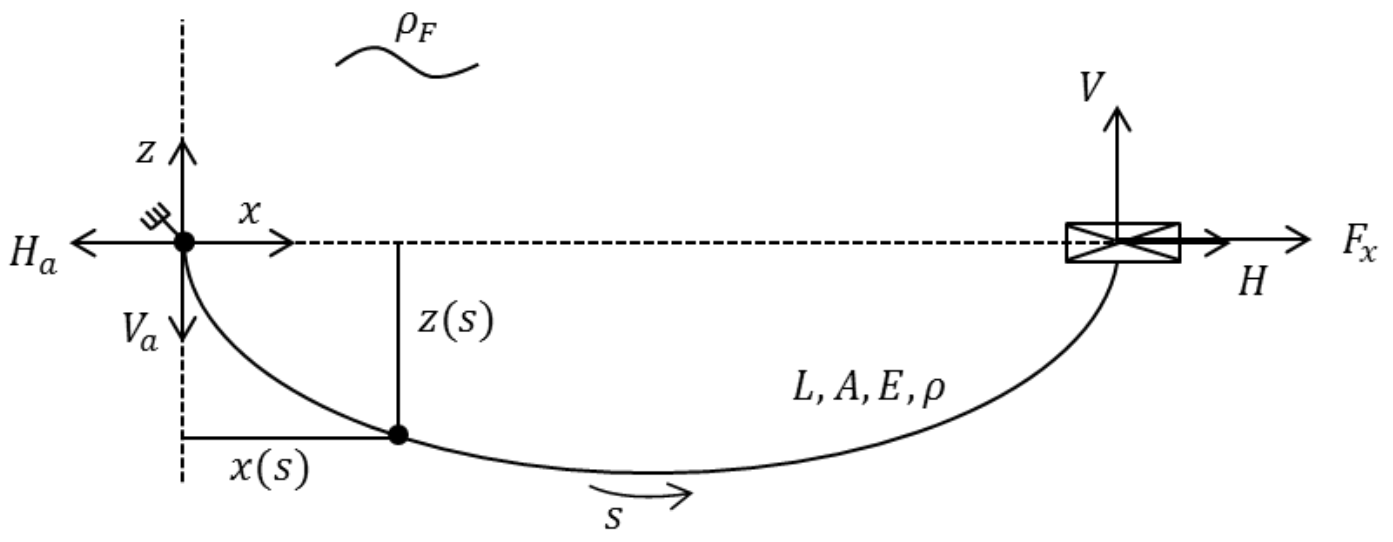

**Figure 9 Example 1: definition of a single catenary segment**

| | |
|---|---|
| Length at rest $L$ | 50 m |
| Young's modulus $E$ | $2.11.10^{11}$ N/m² |
| Area $A$ | $3.1426.10^{-4}$ m² |
| Density $\rho$ | $7.850.10^{3}$ kg/m$^{3}$ |
| Fluid density $\rho_F$ | $1.025.10^{3}$ kg/m$^{3}$ |

**Table 3 Example 1: Segment characteristics**

Figure 10 shows the solutions for values of $c$ ranging from 1 to 10. Table 4 gives the absolute nondimensional errors in positions and maximum tensions along the segment for different values of horizontal tension imposed across the ratio $c = wL/F_x$ with $w = gA(\rho - \rho_F)$. The error is normalized as $\varepsilon_a = \max_{s\in[0,L]} |(n_i - n_{i\,exact})|/wL$ for tension and $\varepsilon_a = \max_{s\in[0,L]} |(x_i - x_{i\,exact})|/L$ for displacements.

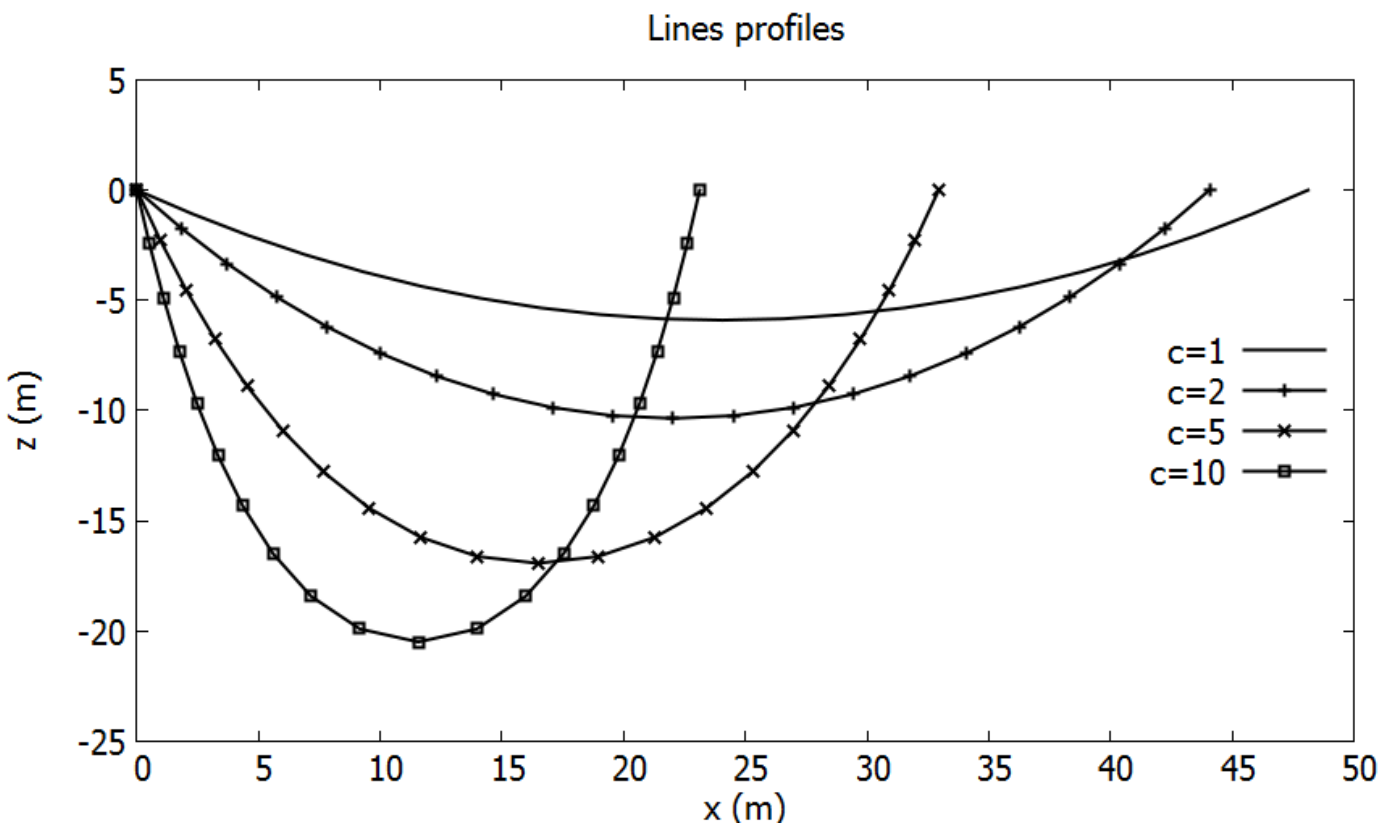

**Figure 10 Example 1: Prismatic ball-joint catenary solutions for different ratios $c$**

| Field | $c = 1$ | $c = 2$ | $c = 5$ | $c = 10$ |
|---|---|---|---|---|
| $x$ | $7.92.10^{-9}$ | $3.79.10^{-8}$ | $5.10.10^{-7}$ | $3.89.10^{-6}$ |
| $z$ | $6.52.10^{-9}$ | $6.61.10^{-8}$ | $8.57.10^{-7}$ | **$7.53.10^{-6}$** |
| $H$ | $1.86.10^{-13}$ | $1.95.10^{-15}$ | $3.73.10^{-14}$ | $1.86.10^{-14}$ |
| $V$ | $4.66.10^{-14}$ | $6.99.10^{-11}$ | $4.84.10^{-14}$ | $4.66.10^{-14}$ |

**Table 4 Example 1: Maximum errors**

The single-shooting method shows very good results in terms of accuracy for these cases. The maximum error for all cases combined does not exceed $7.53.10^{-6}$ for vertical positions with just 10 elements. Errors in the force fields are of the order of the machine precision. This example also enables us to test a mixed boundary condition formulation (prismatic link position and force).

As mentioned in Section 3, an interesting feature of the shooting method is the use of an adaptive step integrator to minimize the number of integration points. Using the RKF45 scheme, we obtain the iterations shown in **Figure 11** (for an error of $10^{-12}$ of the scheme) and the evolution of the number of integration points as a function of the absolute error chosen for the RKF45 scheme in **Figure 12**. These results were obtained for the case $c = 1$ and the initial iteration ${}^{0}\boldsymbol{X} = {}^{0}\boldsymbol{n}_0 = [50, 0, -100]$ (initial condition at the ball joint). Newton's algorithm converges in 5 iterations to the final solution. The initial iteration is not necessarily close to the target (**Figure 11**), we observe a very quick convergence of the method. Moreover, **Figure 12** shows that the number of integration points does not exceed 10 nodes from the first iteration, even for very low absolute scheme errors. Finally, the number of points is identical between iterations 1 and 4. This can be explained by the geometrically close solutions shown in **Figure 11**. An adaptive integration scheme used in conjunction with the shooting method therefore reduces the number of integration nodes and hence the computation time, while maintaining controlled accuracy owing to the integrator's error parameter.

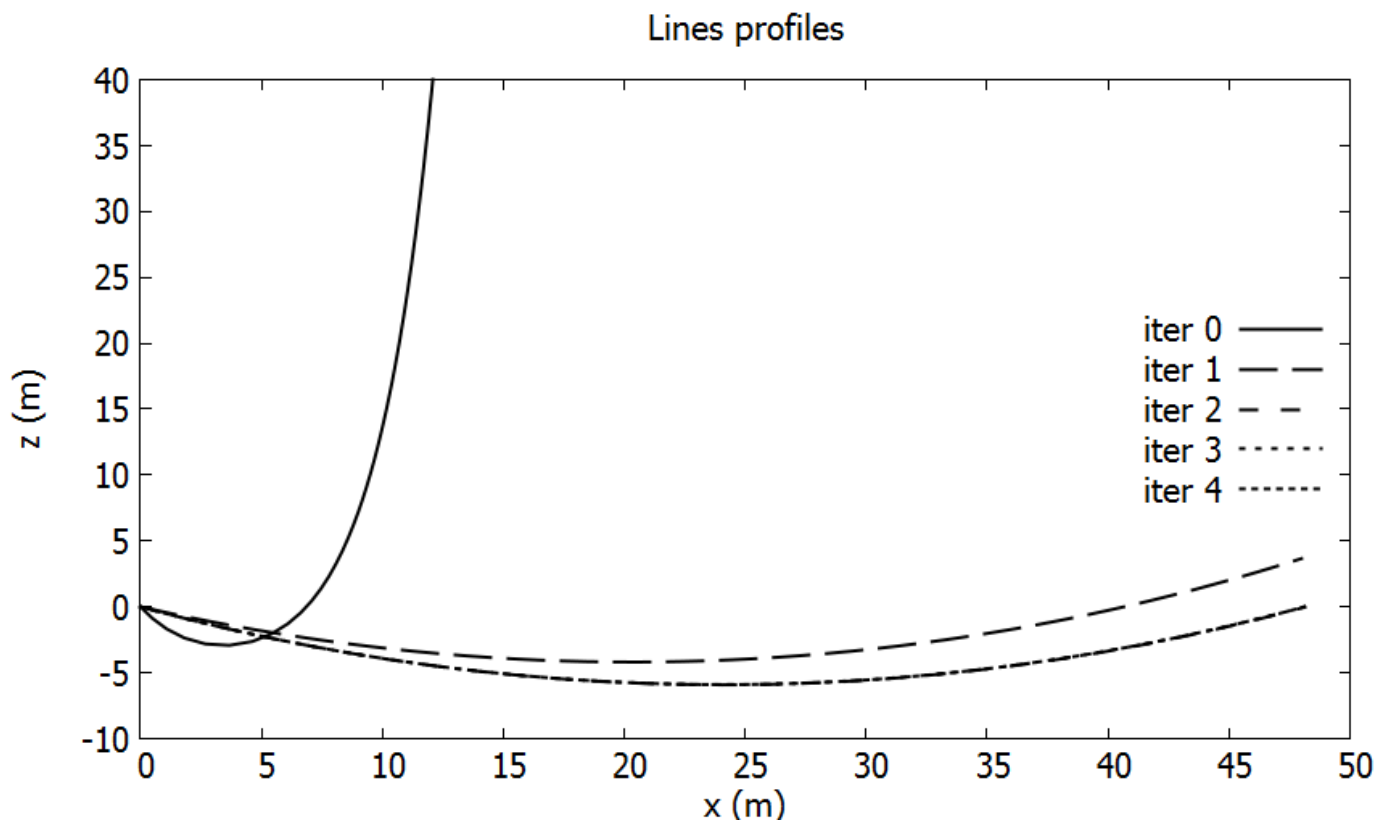

**Figure 11 Example 1: line profiles with regards to Newton iterations**

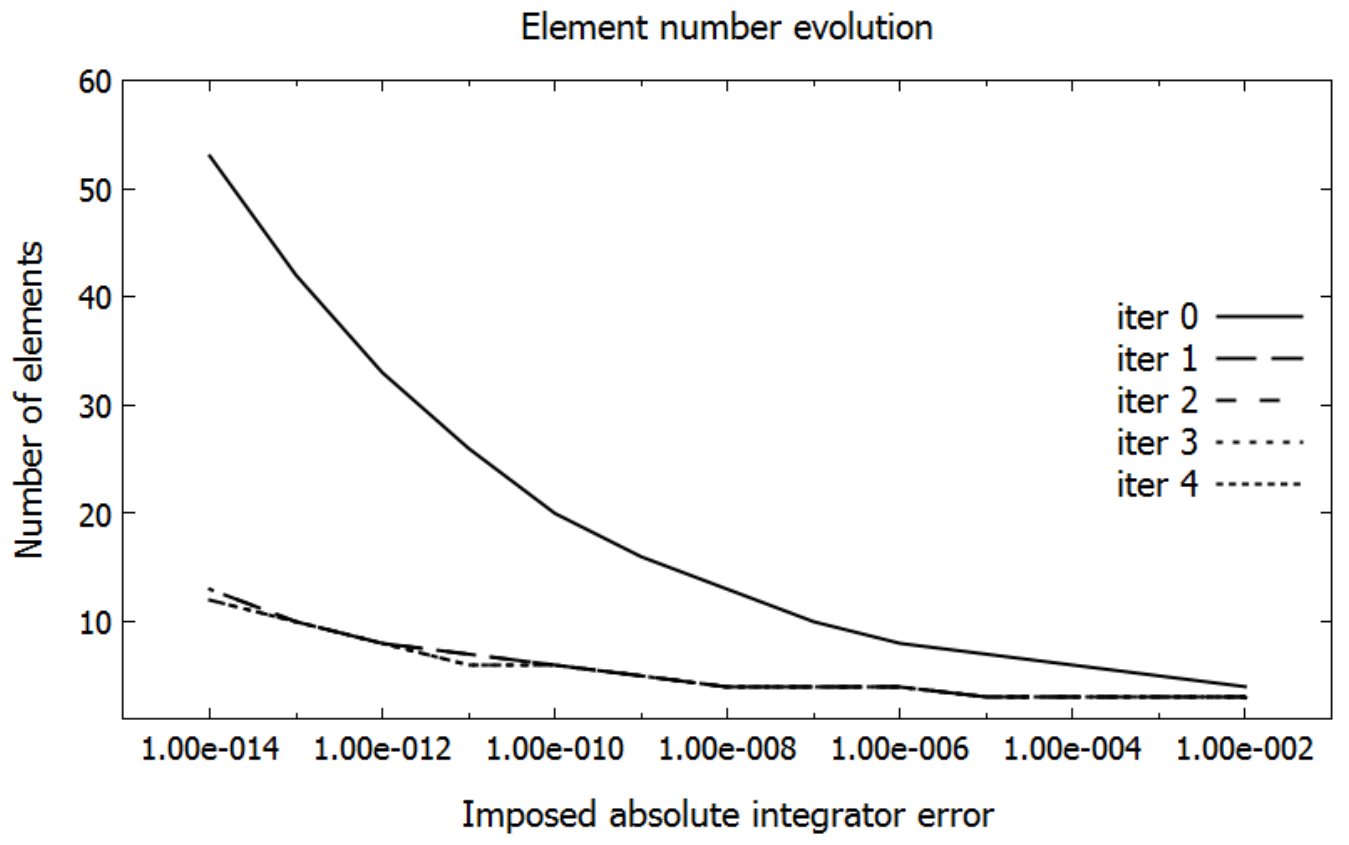

**Figure 12 Example 1: evolution of the number of integration points with regards to the RKF45 scheme imposed absolute error**

### 4.2. Catenary string under several boundary conditions

The aim of this example was to study the behavior of the single shooting method with several types of boundary conditions. A particularity mentioned in Section 3 is the notion of shooting direction. In fact, the ODE is integrated from one side of the simulated segment. Depending on the starting end of the ODE integration, the unknowns and constraints of the shooting method may differ, even if the configuration of the physical system remains the same (Figure 13). The shooting direction also has an impact on the value of the internal force fields, given the positivity convention of right over left, which depends on the origin of the curvilinear abscissas (see Section 2.1). Therefore, we aim to check whether the values of the solution fields are indeed identical, regardless of the choice of the integration direction.

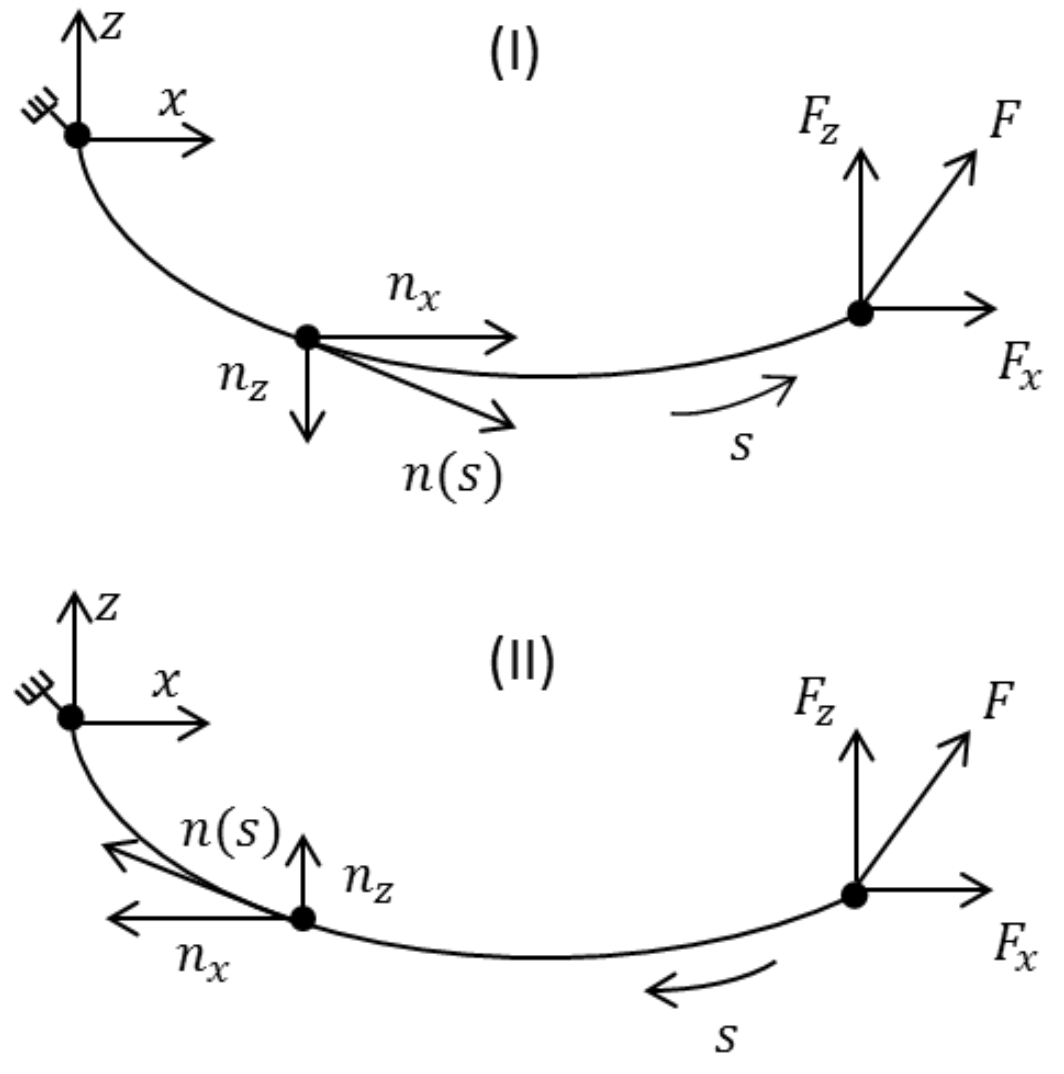

**Figure 13 Example 2: sign convention for forces according to integration direction**

The example is based on a catenary string, whose reference solution is given in Appendix: semi-analytical reference for catenary string examples. The same segment as that in the previous example is considered. It is a simple, deformable segment, with constant material section parameters, following an elastic law, as given in Table 3.

The boundary condition on the left was always considered as a ball joint. The right-hand boundary condition type will be taken in the followings: (a) ball joint, (b) imposed force, (c) prismatic, (e) linear spring, and (f) linear spring +

prismatic link (c.f. Table 1). As the catenary configuration is in a plane, the point connection (d) is equivalent to a prismatic connection for the strings, and is therefore not studied again. These five combinations are duplicated for $s$ increasing and decreasing according to $x$ which correspond to configurations (I) and (II) in Figure 13. Thus, a total of ten combinations were studied in this example.

For case (b), we considered a pure horizontal force such that $F_z = 0$. For cases (b) and (c), we introduce the ratio $c = wL/F_x$ with $w = gA(\rho - \rho_F)$ and such that $F_x$ is obtained for the value $c = 10$. The coordinates of the ball joint at the left end in convention (I) were $x_A = 0$, $z_A = 0$. Finally, the coordinates of the right-hand end for cases (I)-(a), (I)-(c) were $x_F = 25$ and $z_F = 0$. The spring reference point was set at $\boldsymbol{a} = [25, 0,0]^T$. The spring stiffness is $k = wL/10$. The Newton tolerance was set to $10^{-8}$ and the integrator absolute error parameter to $10^{-8}$. The initial iteration estimates for cases (II)-(b), (II)-(c) and (II)-(f) is set to $\boldsymbol{r}_F = [x_F, z_F]$. Similarly, when initial tensions are unknown (cases (I) and (II)-(a), (II)-(c)), vertical and horizontal tensions are assumed to be equal, such that the norm is obtained for the ratio $c = 10$. This corresponds to a string with a tangent that has a downward angle of 45 degrees.

We compare the absolute nondimensional errors $\varepsilon_a = \max_{s\in[0,L]} |(n_i - n_{i\ exact})|/wL$ and $\varepsilon_a = \max_{s\in[0,L]} |(x_i - x_{i\ exact})|/L$ at the beginning, end and maximum along the segment, the results of which are given in Table 5 to Table 7. We removed the zeros from Table 5 because we imposed half of the fields at the initial end. We can see that the maximum error does not exceed $2.92.10^{-9}$ for the position in $x$ position along the segment. At the initial and final ends, the maximum error is $8.29.10^{-10}$ and $2.18.10^{-9}$ respectively, for all fields combined. The errors in case (c) are smaller than those in the previous example because we set the number of integration points constant at 10, whereas in the present example, the RKF45 integrator adapts the step size according to the desired accuracy. Here, we obtain 34 integration points for the final solution of the prismatic ball-and-socket joint case, with $c = 10$, this value having been selected as the least accurate in example 1.

Error values are, therefore, highly dependent on the tolerances chosen for the adaptive step integrator and the Newton algorithm. The tolerance of the Newton algorithm mainly influences the error at the ends, whereas the tolerance of the RKF45 affects the maximum errors along the segment. The adaptive step integrator therefore offers a significant gain in accuracy while limiting the number of nodes to the strict minimum.

This example illustrates that the shooting method can be used to model several types of boundary conditions. Regardless of the configuration chosen (curvilinear abscissa origin), the errors in the position and force fields are of the same order of magnitude, validating the choice of assembling the state vector to be integrated by half the fields known by the boundary condition. The spring-type boundary conditions, considered as classical kinematic links generating the Newton's unknowns and constraints, are also validated. Furthermore, the method is based on the use of continuous fields without any approximation or discretization, which leads to a very high accuracy of the forces along the line, as shown in Table 7.

| | | Fields | | | |
|---|---|---|---|---|---|
| Conv. | Case | $x$ | $z$ | $n_x$ | $n_z$ |
| (I) | (a) | - | - | $2.34.10^{-10}$ | $6.12.10^{-10}$ |
| | (b) | - | - | $1.86.10^{-14}$ | $1.86.10^{-13}$ |
| | (c) | - | - | $1.86.10^{-14}$ | $6.92.10^{-10}$ |
| | (e) | - | - | $2.90.10^{-10}$ | $6.43.10^{-10}$ |
| | (f) | - | - | $2.73.10^{-10}$ | $7.46.10^{-10}$ |
| (II) | (a) | - | - | $2.34.10^{-10}$ | $6.12.10^{-10}$ |
| | (b) | $7.60.10^{-10}$ | $3.03.10^{-11}$ | - | - |
| | (c) | **$8.29.10^{-10}$** | - | $1.86.10^{-14}$ | $6.92.10^{-10}$ |
| | (e) | $4.10.10^{-11}$ | $6.80.10^{-12}$ | $2.39.10^{-10}$ | $5.72.10^{-10}$ |
| | (f) | $4.79.10^{-11}$ | - | $2.73.10^{-10}$ | $7.46.10^{-10}$ |

**Table 5 Example 2: Absolute nondimensional errors at the initial endpoint**

| | | Fields | | | |
|---|---|---|---|---|---|
| Conv. | Case | $x$ | $z$ | $n_x$ | $n_z$ |
| (I) | (a) | $2.00.10^{-13}$ | $4.51.10^{-15}$ | $2.34.10^{-10}$ | $6.12.10^{-10}$ |
| | (b) | **$2.18.10^{-9}$** | $1.13.10^{-10}$ | $1.86.10^{-14}$ | $1.90.10^{-15}$ |
| | (c) | $8.29.10^{-10}$ | $4.77.10^{-16}$ | $1.86.10^{-14}$ | $6.92.10^{-10}$ |
| | (e) | $5.06.10^{-11}$ | $4.78.10^{-12}$ | $2.90.10^{-10}$ | $6.43.10^{-10}$ |
| | (f) | $4.79.10^{-11}$ | $2.53.10^{-15}$ | $2.73.10^{-10}$ | $7.46.10^{-10}$ |
| (II) | (a) | $1.04.10^{-13}$ | $2.88.10^{-15}$ | $2.34.10^{-10}$ | $6.12.10^{-10}$ |
| | (b) | $1.36.10^{-15}$ | $2.93.10^{-16}$ | $1.86.10^{-14}$ | $1.86.10^{-13}$ |
| | (c) | $4.22.10^{-15}$ | $1.71.10^{-15}$ | $1.86.10^{-14}$ | $6.92.10^{-10}$ |
| | (e) | $7.97.10^{-16}$ | $1.06.10^{-16}$ | $2.39.10^{-10}$ | $5.72.10^{-10}$ |
| | (f) | $1.78.10^{-15}$ | $6.90.10^{-17}$ | $2.73.10^{-10}$ | $7.46.10^{-10}$ |

**Table 6 Example 2: Absolute nondimensional errors at the end point**

| | | Fields |
|---|---|---|

| Conv. | Case | $x$ | $z$ | $n_x$ | $n_z$ |
|---|---|---|---|---|---|
| | (a) | $8.83.10^{-10}$ | $1.10.10^{-10}$ | $2.34.10^{-10}$ | $6.12.10^{-10}$ |
| | (b) | **$2.92.10^{-9}$** | $1.59.10^{-10}$ | $1.86.10^{-14}$ | $4.42.10^{-13}$ |
| (I) | (c) | $1.26.10^{-9}$ | $1.20.10^{-10}$ | $1.86.10^{-14}$ | $6.92.10^{-10}$ |
| | (e) | $1.09.10^{-9}$ | $1.16.10^{-10}$ | $2.90.10^{-10}$ | $6.43.10^{-10}$ |
| | (f) | $9.47.10^{-10}$ | $1.13.10^{-10}$ | $2.73.10^{-10}$ | $7.46.10^{-10}$ |
| | (a) | $8.83.10^{-10}$ | $1.10.10^{-10}$ | $2.34.10^{-10}$ | $6.12.10^{-10}$ |
| | (b) | $1.18.10^{-9}$ | $6.03.10^{-11}$ | $1.86.10^{-14}$ | $1.86.10^{-13}$ |
| (II) | (c) | $1.20.10^{-9}$ | $1.20.10^{-10}$ | $1.86.10^{-14}$ | $6.92.10^{-10}$ |
| | (e) | $9.54.10^{-10}$ | $9.66.10^{-11}$ | $2.39.10^{-10}$ | $5.72.10^{-10}$ |
| | (f) | $9.55.10^{-10}$ | $1.13.10^{-10}$ | $2.73.10^{-10}$ | $7.46.10^{-10}$ |

**Table 7 Example 2: Maximum absolute nondimensional errors along the section**

### 4.3. Example 3: string subjected to a following radial force

In the two previous examples, we have seen the application of the shooting method to catenary cases with several types of boundary conditions. The catenary equations are based on a gravity force that does not depend on the position and orientation of the section. This leads to good conditioning of the IVP, as long as the line is not tensioned.

We are now testing a simple shooting method applied to the resolution of a section subjected to a position-dependent constraint. This type of approach was previously used in the offshore field for anchor lines subjected to the effects of gravity and current by [2], [3] with a formalism of string kinematics in spherical coordinates.

Here, we will validate the Cartesian formalism adopted by carrying out an example study on the Velaria reference [16]. The string is assumed to be inextensible, massless, and subjected to a normal distributed force of constant intensity $p$ of the form $\boldsymbol{f}(s) = -p\boldsymbol{k} \wedge \partial_s \boldsymbol{r}$. The solution to this problem is a circular arc. This example corresponds to the application of a hydrostatic pressure of amplitude $p$ on a cylindrical membrane of axis $\boldsymbol{k}$. Here, we consider an inextensible string (constitutive law (10)) whose ends are ball joints (Figure 14). The radius $R$ was set to 1 m, and the distributed force $p$ was 1 N/m. The ball joint boundary conditions impose $\boldsymbol{r}(s = 0) = R.\boldsymbol{x}$ and $\boldsymbol{r}(s = L) = -R.\boldsymbol{x}$. With this parameterization, we have the segment length $L = \pi R$, and the curvilinear abscissa origin is chosen for an angle $\theta$ such that $\theta(s) = \frac{s}{R}$.

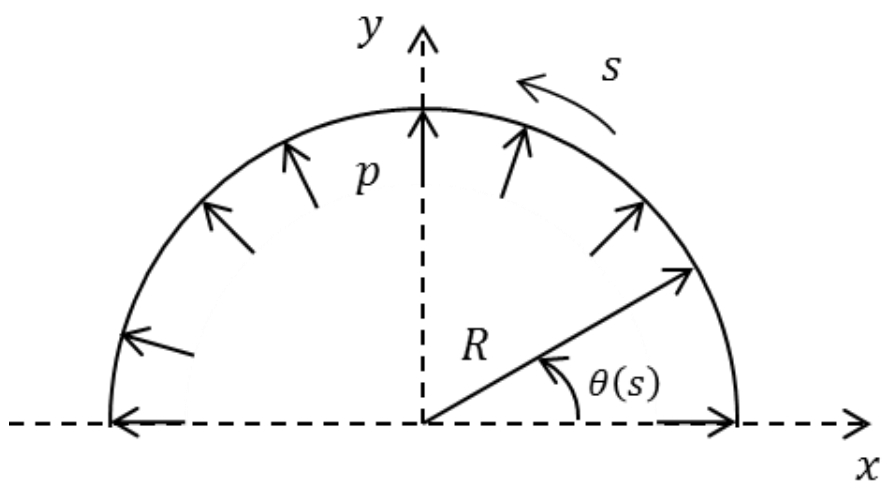

**Figure 14 Example 3: Constant hydrostatic pressure on a cylindrical membrane**

The analytical solution for this problem is [16]:

$$x(s) = Rcos\big(\theta(s)\big) \tag{47}$$

$$y(s) = Rsin\big(\theta(s)\big)$$

$$n_x(s) = -pRsin\big(\theta(s)\big)$$

$$n_y(s) = pRcos\big(\theta(s)\big)$$

Even though the shooting algorithm is numerically solved in three dimensions, we carry out reasoning in the plane for the analytical solution. Conceptually, this case is 2-dimensional, with two unknowns for the Newton algorithm ($X = \big[n_x(s=0), n_y(s=0)\big]^T$).

We introduce the dimensionless tension $\lambda$ (percentage of solution tension) at the starting end, defined as $n_y(s=0) = \lambda pR = n_y^\lambda$ such that $X^\lambda = \big[0, n_y^\lambda\big]^T$ (the $n_x(s=0)$ component is cancelled out to simplify the analysis).

The Newton tolerance is taken to be $10^{-10}$, the absolute error for the integrator is $10^{-12}$. Two initial estimates are investigated: $\lambda = -1.5$ and $\lambda = 1.5$. The shooting method converges in 9 iterations for each case (Figure 15).

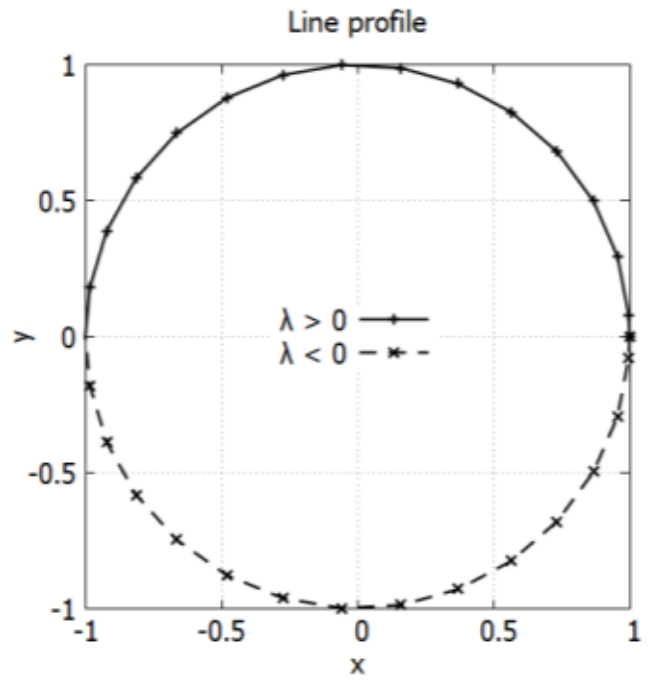

**Figure 15 Example 3: multiple solutions of the pull method for a string subjected to a radial force**

If we parameterize the final constraint equation $\boldsymbol{r}(s = L) = -R.\boldsymbol{x}$ by $\lambda$ we obtain two terms: $C_x(\lambda) = x(s = L; \lambda)/R + 1$ and $C_y(\lambda) = y(s = L; \lambda)/R$. Figure 16 shows the evolution of $C_x(\lambda)$ and $C_y(\lambda)$ as functions of $\lambda$. Because the aim of the shooting method is to cancel these functions through the Newton, the solution is obtained for $\lambda = 1$. The phase diagram $\left(C_x(\lambda), C_y(\lambda)\right)$ is shown in Figure 17, where the labels indicate the values taken by $\lambda$ to initialize $X^\lambda$. In this phase diagram, the solution is found for $\left(C_x(\lambda), C_y(\lambda)\right) = (0{,}0)$. The solid line corresponds to the case $\boldsymbol{k} = \boldsymbol{z}$ with $\lambda > 0$. Figure 16 and Figure 17 are simply obtained by integrating ODE for several values of $\lambda \in [-1.7, -0.5] \cup [0.5{,}1.7]$ and therefore $X^\lambda$.

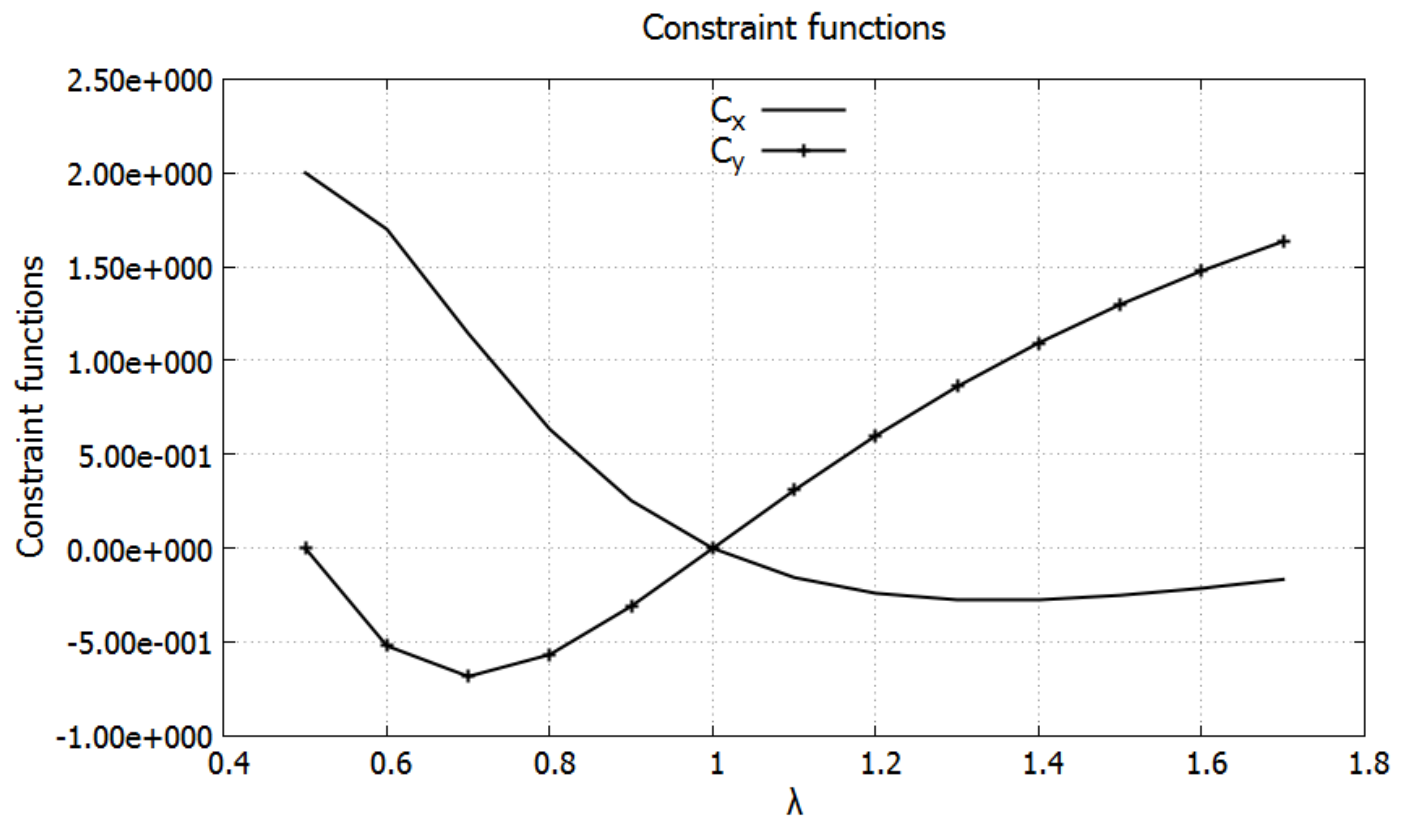

**Figure 16 Example 3: Changes in $C_x(\lambda)$ and$C_y(\lambda)$ for a radial linear force**

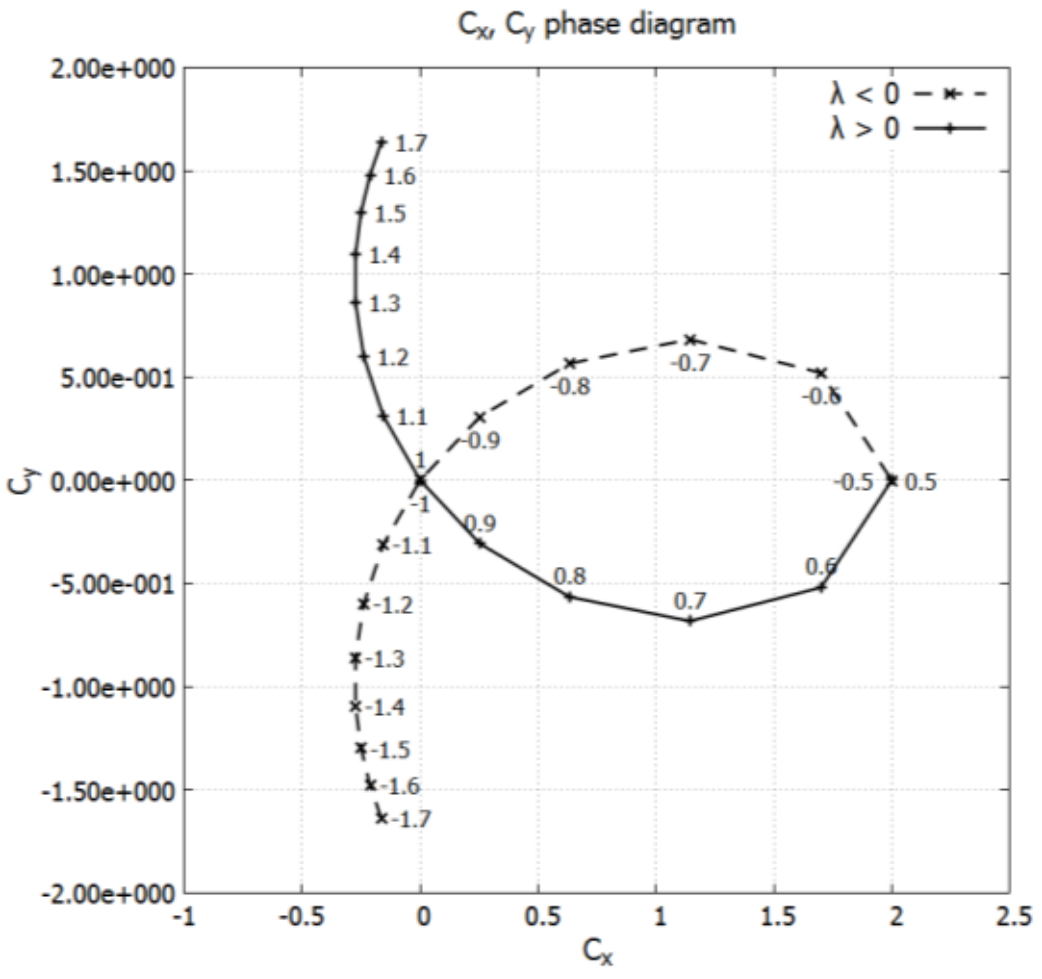

**Figure 17 Example 3: phase diagram $C_x(\lambda)$ and$C_y(\lambda)$ function of values taken by $\lambda$**

If we modify the normal $\boldsymbol{k}$ such that $\boldsymbol{k} = \frac{\boldsymbol{OM} \wedge \partial_s \boldsymbol{r}}{\|\boldsymbol{OM} \wedge \partial_s \boldsymbol{r}\|}$ where $M$ is a current point on the segment, the parameterization can be extended to handle either hourly or trigonometric solutions of the final position of the segment (direction of angle $\theta$ and curvilinear abscissa). Thus, the dotted solution in Figure 15 can be reached, and two solutions symmetrical to $\boldsymbol{x}$ exist ($\lambda$ positive and $\lambda$ negative). The dotted line in the phase diagram is obtained for negative values of $\lambda$.

It is interesting to note from Figure 17 that, for such a simple two-dimensional example, where the evolution of the constraint functions is restricted to the single parameter $\lambda$, two solutions can coincide when the phase diagram intersects for different values of $\lambda$. Each solution can be reached by the shooting method as long as the initial iteration lies within the convergence domain of the solution.

If we look closely at Figure 17, we can see another point of intersection $\left(C_x(\lambda), C_y(\lambda)\right) = (2,0)$ for the values $\lambda = -0.5$ and $\lambda = 0.5$. This corresponds to the case of the closed circle. Because the reference configuration is an open segment (disconnected ends), there are an infinite number of solutions for this specific point on the phase diagram, corresponding to full circles with radii $r = \frac{R}{n}$ (the string turns on itself).

Finally, the maximum nondimensional errors along the segment $\varepsilon_a = \max_{s \in [0,L]} |(n_i - n_{i\,exact})|/pR$ and $\varepsilon_a = \max_{s \in [0,L]} |(x_i - x_{i\,exact})|/R$ are given in Table 8.

| Fields | Initial iteration $\lambda$ | |
|---|---|---|
| | -1.5 | 1.5 |
| $x$ | $3.70.10^{-9}$ | $3.70.10^{-9}$ |
| $y$ | $9.18.10^{-9}$ | $9.18.10^{-9}$ |
| $n_x$ | $6.01.10^{-9}$ | $6.01.10^{-9}$ |
| $n_y$ | $4.58.10^{-9}$ | $4.58.10^{-9}$ |

**Table 8 Example 3: Maximum nondimensional errors**

Once again, the precision of the shooting method is remarkable in this example, where the linear force depends on the geometry of the segment in terms of position and orientation.

Here, we highlight the fact that the shooting method can take into account nonlinear loading on the slender body, an inextensible material law, and reach several solutions of the problem at both ends when they exist (provided we have a sufficiently well-defined first iteration) while retaining quasi-analytical accuracy.

### 4.4. Example 4: three segments converging strings configuration

This example illustrates the use of mixed the multi-body/multi-shooting method to solve the problem of several catenary segments (anchor lines) connected together, as shown in Figure 18 and Figure 19.

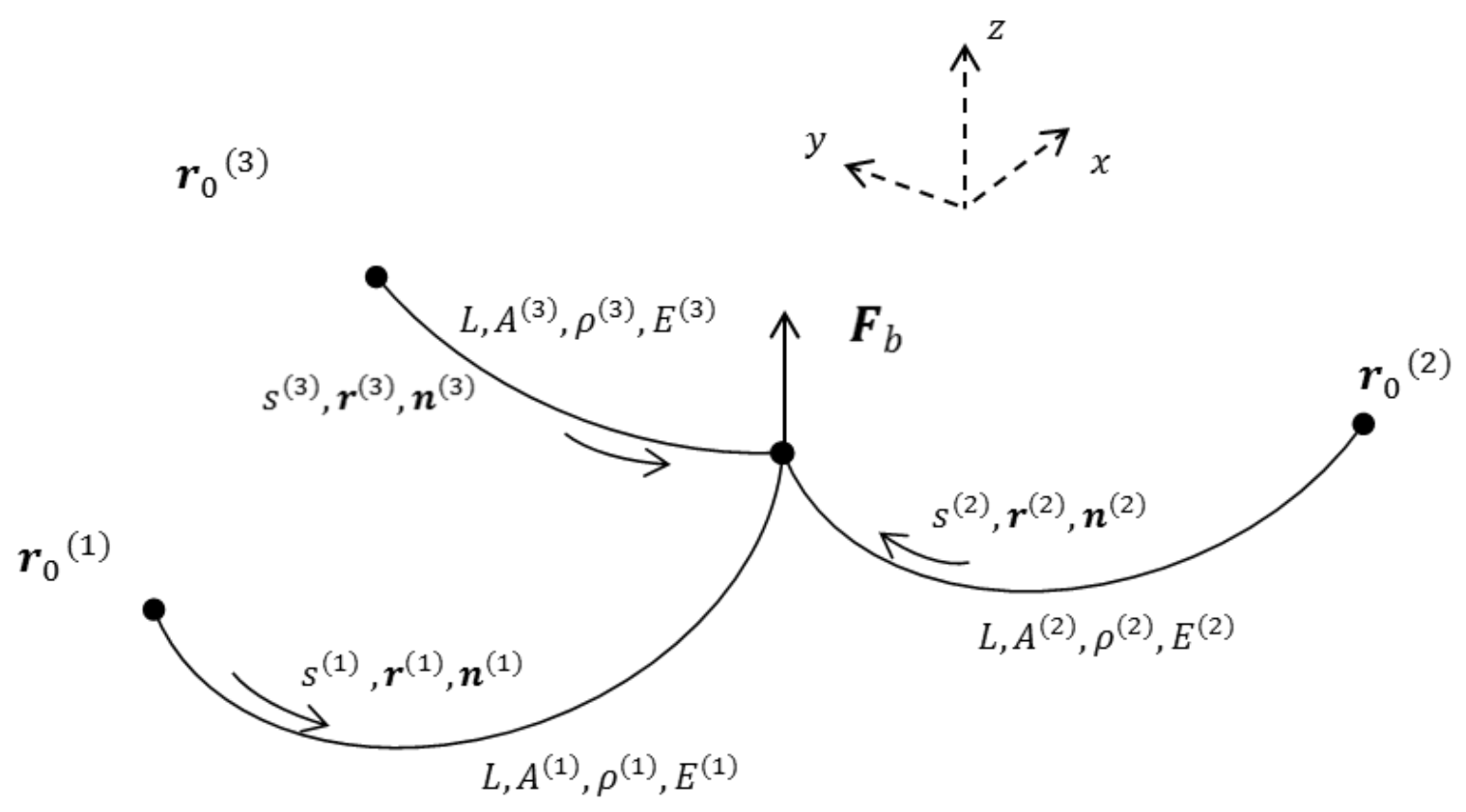

**Figure 18 Example 4: configuration of the three-segment example - 3D view**

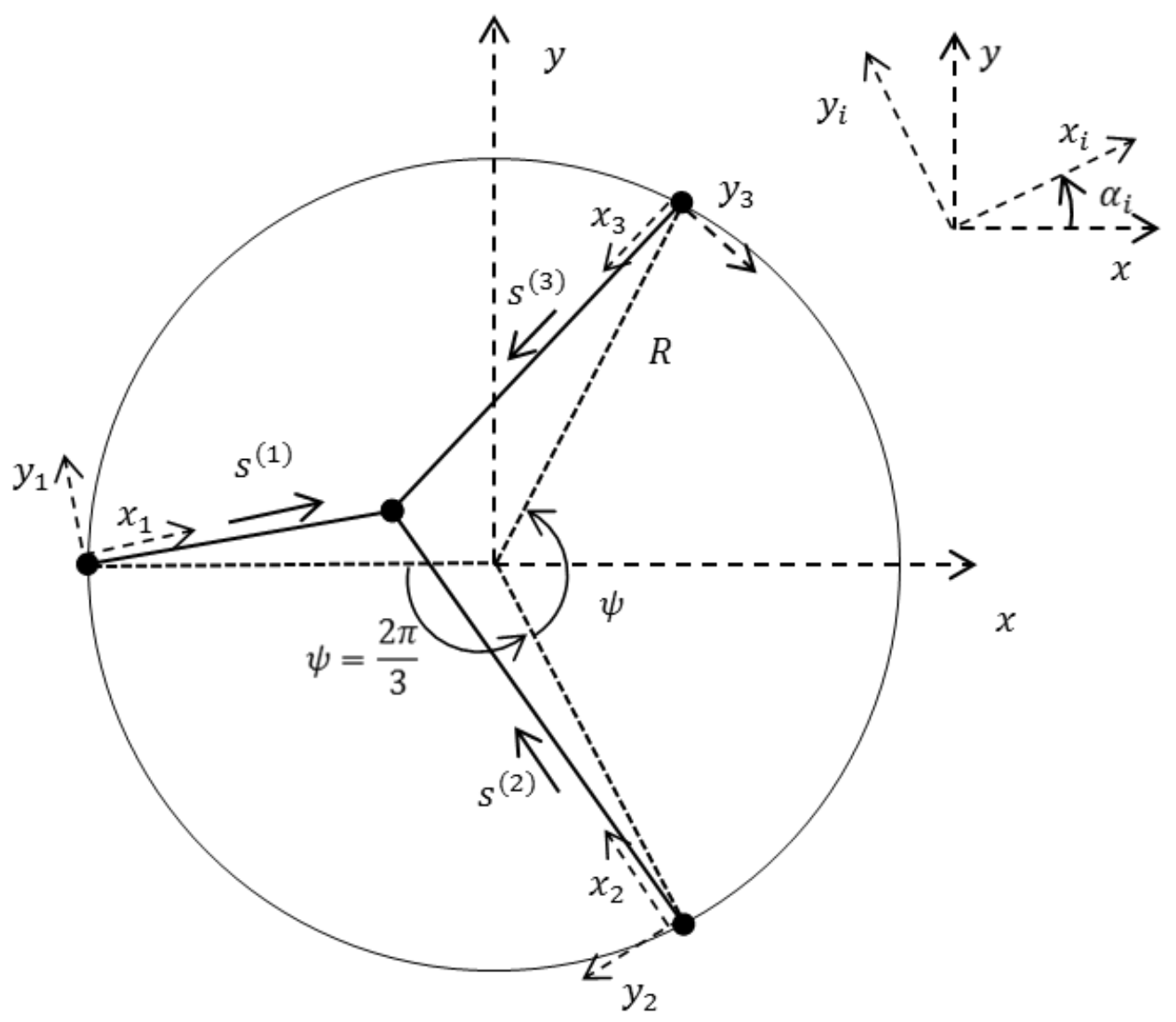

**Figure 19 Example 4: configuration of the three-segment example - horizontal view**

We consider three extremities uniformly distributed on a circle of radius $R$ at constant angles of $\psi = \frac{2\pi}{3}$ and zero altitude. Each of the three segments has its own material characteristics $E^{(i)}, A^{(i)}, \rho^{(i)}$and they are connected together at one end by a buoy modeled by a vertical force $\boldsymbol{F}_b$. The semi-analytical resolution from the catenary equations is given in Appendix: semi-analytical reference for catenary string examples.

We consider the following values: $L_0 = 50\ m$, $E_0 = 2.11.10^{11}\ N/m^2, \rho_0 = 7850\ kg/m^3$ and $A_0 = 3.1416.10^{-4}\ m^2$. Based on these parameters, the properties of each segment are given in Table 9.

| Property | Segment 1 | Segment 2 | Segment 3 |
|---|---|---|---|
| $L^{(i)}$ | $L_0$ | $L_0$ | $L_0$ |
| $A^{(i)}$ | $A_0$ | $A_0$ | $A_0$ |
| $E^{(i)}$ | $0.05E_0$ | $0.001E_0$ | $E_0$ |
| $\rho^{(i)}$ | $2\rho_0$ | $0.5\ \rho_0$ | $\rho_0$ |

**Table 9 Example 4: Segment properties**

For this example, a tolerance of $10^{-10}$ was used for the Newton algorithm and $10^{-12}$ for the absolute error of the adaptive step integrator. The initial iteration is taken at for each segment at ${}^0X^i = {}^0X cos(\psi_i)$ with ${}^0X = [50, 0, -100]$.

Figure 20 shows the final positions of each segment in the horizontal plane. The maximum absolute nondimensional errors $\varepsilon_a = \max_{s\in[0,L]} |(n_i - n_{i\ exact})|/wL$ for tensions and $\varepsilon_a = \max_{s\in[0,L]} |(x_i - x_{i\ exact})|/L$ for the positions are presented for each segment in Table 10. Line angles, buoy position, and associated dimensional absolute errors ($\varepsilon_a = |(X - X_{exact})|$) are given in Table 11. In addition, Table 12 provides the number of endpoints used by the adaptive step integrator to integrate the ODE of each section.

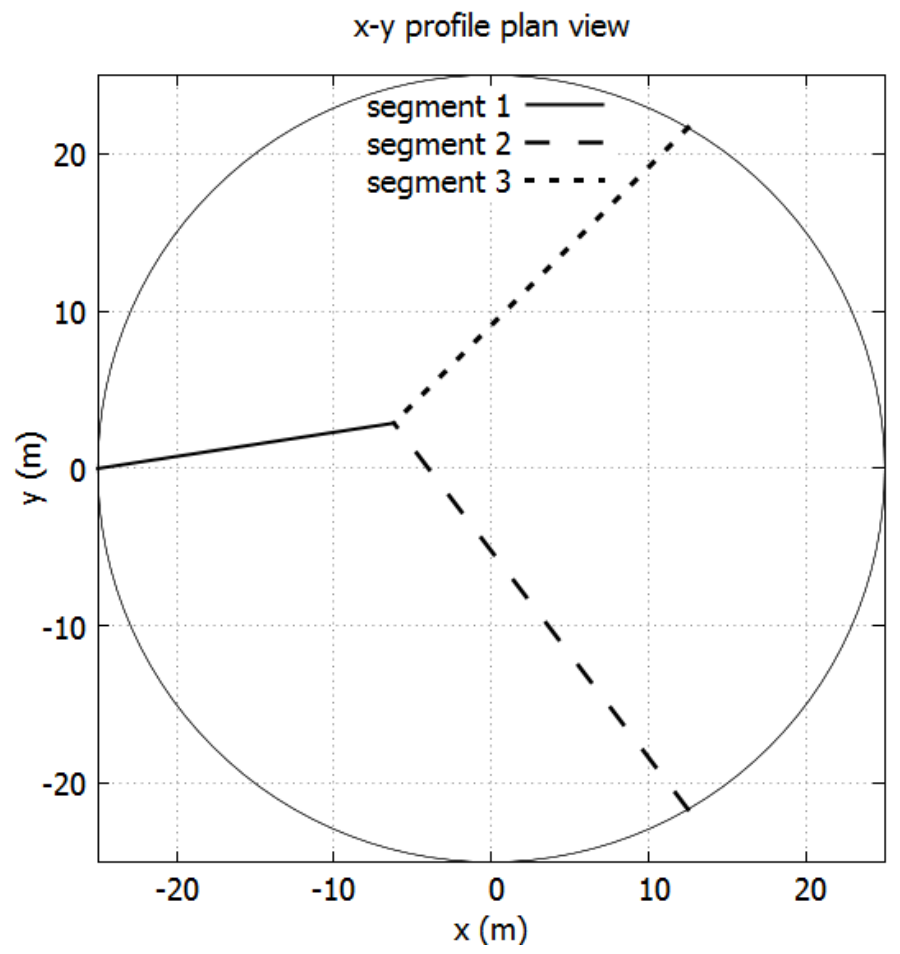

**Figure 20 Example 4: Final positions of segments in the x-y plane**

| Fields | Line 1 | Line 2 | Line 3 |
|---|---|---|---|
| $x$ | $1.67.10^{-11}$ | $2.09.10^{-11}$ | $1.88.10^{-11}$ |
| $y$ | $1.11.10^{-11}$ | $1.08.10^{-10}$ | $1.08.10^{-10}$ |
| $z$ | $5.19.10^{-12}$ | $3.24.10^{-11}$ | $1.87.10^{-11}$ |
| $H$ | $5.85.10^{-12}$ | $5.16.10^{-11}$ | $2.68.10^{-11}$ |
| $V$ | $8.97.10^{-13}$ | $2.60.10^{-12}$ | $4.10.10^{-13}$ |

**Table 10 Example 4: Maximum nondimensional errors for each section**

| Fields | Multi-body /multi-shoot | Reference | Absolute error (dimensional) |
|---|---|---|---|
| $\alpha^{(1)}$ (rad) | $1.5208.10^{-1}$ | $1.5208.10^{-1}$ | $2.2130.10^{-11}$ |
| $\alpha^{(2)}$ (rad) | 2.2216 | 2.2216 | $1.3838.10^{-10}$ |
| $\alpha^{(3)}$ (rad) | -2.3540 | -2.3540 | $1.5105.10^{-10}$ |
| $x_b$ (m) | -6.1827 | -6.1827 | $8.3703.10^{-10}$ |
| $y_b$ (m) | 2.8840 | 2.8840 | $5.5509.10^{-10}$ |
| $z_b$ (m) | 8.8387 | 8.8387 | $9.1566.10^{-11}$ |

**Table 11 Example 4: Results and errors for main parameters**

| | Segment 1 | Segment 2 | Segment 3 |
|---|---|---|---|
| Elements | 33 | 24 | 27 |

**Table 12 Example 4: Number of elements from the adaptive step integrator**

The results in Table 10 and Table 12 show very good accuracy of the multiple shooting method on the fields $\boldsymbol{r}$ and $\boldsymbol{n}$ fields, line angles and final buoy position. This result is not surprising because the catenary resolution method is a derivative of the shooting method. Indeed, we can draw a parallel between the shooting method and the catenary equations, for which the equations (51) to (54) of Appendix: semi-analytical reference for catenary string examples are the result of the analytical integration of the ODE. The algebraic equations (51) to (54) associated with the boundary condition values are solved using a Newton algorithm. This solving process for catenaries is only possible thanks to the form of the differential equation derived from the equilibrium equations, which only take account of gravity as an external force. Analytical integration of forces is not possible in the general case, such as a current force [2], [3].

In addition to the precision aspects, this example validates the mixed multi-body/multi-shooting approach in which the buoy is modeled by a rigid body to which each line is connected (in a ball-and-socket joint) and to which the force $\boldsymbol{F}_b$ is applied.

The adaptive integrator also enables automatic adjustment of the integration step to target the required tolerance. We can see a trend here: the number of points increases with the linear mass, i.e., with the external force of gravity imposed, which induces greater deformation. The same observation was made between examples n°1 and n°2. The trend is less marked for variations in the Young's modulus.

## 5. Conclusion

In this paper, we proposed a new multi-body/multi-shooting method compatible with a multi-body simulation context for the study of discontinuities (geometric, material, or force) and 3D static slender body assemblies of strings used in the offshore industry. This new method stems from the formalism of the classical single and multiple shooting methods. It is positioned as an extension of the multiple-shooting method by introducing static rigid bodies on which slender bodies are connected via kinematic links. The introduction of this static body concept into the shooting method bridges the gap between classical static multi-body resolutions, where rigid bodies are directly connected to each other via kinematic links.

The set of boundary conditions for strings was reviewed, as well as their place within the shooting method, in the form of constraint equations based on the kinematics and permissible statics of each connection.

We have observed that the use of an adaptive step integration algorithm enables us to maintain an optimum ratio between the number of integration points and accuracy. On the one hand, this eliminates the need for a convergence study of the integrated fields, while maintaining optimum computation times.

A Newton algorithm was used to determine the zeros of the constraint function of the shooting problem, allowing quadratic convergence to the solution. We showed that, compared with the single-shooting method, the multiple-shooting method asymptotically tends to a linear solution on each sub-segment when used to solve a single slender body. The division of the integration interval into sub-intervals brings the problem closer to linear at the cost of a significant increase of unknowns.

The multi-body/multi-shooting method has been derived from the classical multi-shooting method by splitting the continuity equation at cuts owing to the introduction of a rigid body on which a slender body is connected. We demonstrated that, contrary to the increase in unknowns between the single shooting method and the multi-shooting method, there is no increase in unknowns between the multi-shooting method and the multi-body/multi-shooting method. The latter also has the advantage of generalizing application cases compared with the multi-shooting method.

The multi-body/multi-shooting method was then validated on four reference cases from the offshore industry. The numerical results showed remarkable accuracy for every configuration tested: every combination of starting/ending boundary conditions, catenary or nonlinear position-dependent lineic forces, and assemblies. These results demonstrate the versatility and capacity of the multi-body/multi-shooting method for simulating 3D nonlinear static strings.

## 7. Appendix: semi-analytical reference for catenary string examples

Differential equations for catenary statics

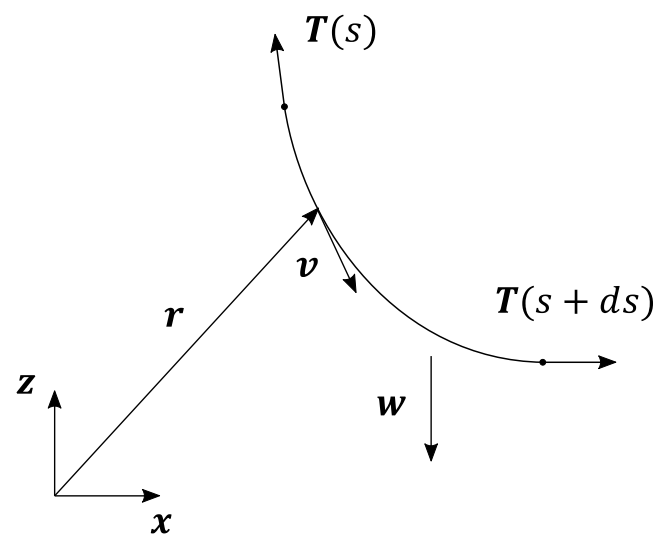

**Figure 21 Balance of a catenary section**

Solving the catenary of an elastic string follows an algebraic equation solving process [21] which is recalled here. Consider a line of undeformed length $L$ following string kinematics, subject only to its weight. Let's note $\boldsymbol{r}(s) = [x(s), z(s)]^T$ the position of a material section and $\boldsymbol{T}(s) = [H(s), V(s)]^T$ the tension vector. In addition, we denote the values at the ends by:

$$\boldsymbol{r}(s=0) = [x_0, z_0]^T$$
$$\boldsymbol{T}(s=0) = [H_0, V_0]^T$$

$$\boldsymbol{r}(s=L) = [x_L, z_L]^T$$
$$\boldsymbol{T}(s=L) = [V_L, V_L]^T$$

We introduce $w = gA\rho$ with $g$ the acceleration of gravity, $A$ the material cross-sectional area and $\rho$ the density (of the material $\rho_{mat}$ or relative to the mass of a fluid $(\rho_{mat} - \rho_{fluid})$). In this case, the balance of forces on the cable is expressed as follows:

$$\boldsymbol{T}(s+ds)-\boldsymbol{T}(s)-w\boldsymbol{z}=\boldsymbol{0}$$

$$\frac{\partial \boldsymbol{T}(s)}{\partial s}-w\boldsymbol{z}=\boldsymbol{0}$$

Let $\partial_s \boldsymbol{r}(s)=\boldsymbol{v}(s)=\left[\frac{\partial x(s)}{\partial s},\frac{\partial z(s)}{\partial s}\right]^T$ be the tangent vector to the curve and $v(s)=|\boldsymbol{v}(s)|=\sqrt{\frac{\partial x(s)}{\partial s}^2+\frac{\partial z(s)}{\partial s}^2}$ then:

$$\frac{\partial}{\partial s}\left(T(s)\frac{\boldsymbol{v}(s)}{v(s)}\right)-w\boldsymbol{z}=\boldsymbol{0}$$

Either by projecting onto the axes $\boldsymbol{x}$ and $\boldsymbol{z}$:

$$\begin{cases}\frac{\partial H(s)}{\partial s}=\frac{\partial}{\partial s}\left(T(s)\frac{1}{v(s)}\frac{\partial x(s)}{\partial s}\right)=0\\ \frac{\partial V(s)}{\partial s}=\frac{\partial}{\partial s}\left(T(s)\frac{1}{v(s)}\frac{\partial z(s)}{\partial s}\right)=w\end{cases} \tag{48}$$

If we consider the tension boundary conditions $\boldsymbol{T}(s=0).\boldsymbol{x}=H_0$ and $\boldsymbol{T}(s=0).\boldsymbol{z}=V_0$ then:

$$\begin{cases}H(s)=H_0\\ V(s)=V_0+ws\end{cases} \tag{49}$$

$$\begin{cases}T(s)\frac{\partial x(s)}{\partial s}=H_0 v(s)\\ T(s)\frac{\partial z(s)}{\partial s}=(V_0+ws)v(s)\end{cases}$$

Using the last two equations, we obtain:

$$T(s)=\sqrt{H_0^2+(V_0+ws)^2}$$

The material law yields the following:

$$T(s)=EA(v(s)-1)$$

Or:

$$v(s)=1+\frac{T(s)}{EA}$$

Thus, by replacing $T(s)$ and $v(s)$ with their expressions we obtain the following differential equations:

$$\begin{cases} T(s)\dfrac{\partial x(s)}{\partial s} = H_0 v(s) \\ T(s)\dfrac{\partial z(s)}{\partial s} = (V_0 + ws)v(s) \end{cases}$$

$$\begin{cases} \dfrac{\partial x(s)}{\partial s} = \dfrac{1}{\sqrt{1+\left(\dfrac{V_0 + ws}{H_0}\right)^2}} + \dfrac{H_0}{EA} \\ \dfrac{\partial z(s)}{\partial s} = \dfrac{(V_0 + ws)}{\sqrt{H_0^2 + (V_0 + ws)^2}} + \dfrac{(V_0 + ws)}{EA} \end{cases}$$

Their integration yields the following equation:

$$\begin{cases} x(s) = \dfrac{H_0}{w}\left[\sinh^{-1}\left(\dfrac{V_0 + ws}{H_0}\right) - \sinh^{-1}\left(\dfrac{V_0}{H_0}\right)\right] + \dfrac{H_0 s}{EA} + x_0 \\ z(s) = \dfrac{H_0}{w}\left[\sqrt{1+\left(\dfrac{V_0 + ws}{H_0}\right)^2} - \sqrt{1+\left(\dfrac{V_0}{H_0}\right)^2}\right] + \dfrac{ws}{EA}\left(\dfrac{V_0}{w} - \dfrac{s}{2}\right) + z_0 \end{cases} \quad (50)$$

To obtain (49) and (50), we only used the boundary conditions at $s = 0$. Evaluating these equations in $s = L$ we obtain:

$$x_L = \frac{H_0}{w}\left[\sinh^{-1}\left(\frac{V_0 + wL}{H_0}\right) - \sinh^{-1}\left(\frac{V_0}{H_0}\right)\right] + \frac{H_0 L}{EA} + x_0 \quad (51)$$

$$z_L = \frac{H_0}{w}\left[\sqrt{1+\left(\frac{V_0 + wL}{H_0}\right)^2} - \sqrt{1+\left(\frac{V_0}{H_0}\right)^2}\right] + \frac{wL}{EA}\left(\frac{V_0}{w} - \frac{L}{2}\right) + z_0 \quad (52)$$

$$H_L = H_0 \quad (53)$$

$$V_L = V_0 + wL \quad (54)$$

System (51) to (54) is the result of the analytical integration of the differential system governing the statics of a string under its weight. This constitutes a system of algebraic equations to be solved. The boundary conditions in $s = 0$ and $s = L$ we'll provide a quadruplet among the values of $x_0$, $z_0$, $x_L$, $z_L$, $H_0$, $V_0$, $H_L$ and $V_L$. For example, when designing an anchor line, the end positions are given and the forces are determined. In the case of an imposed force in $s = L$ the parameters $x_0$, $z_0$, $H_L$ and $V_L$ will be supplied, enabling the constraints to be written as a function of the unknowns $x_L$, $z_L$, $H_0$ and $V_0$.